\documentclass[aps, prb, twocolumn,amsmath,amssymb,floatfix,reprint,longbibliography]{revtex4-2}
\usepackage{graphicx}
\usepackage{physics}
\usepackage{times}
\usepackage{dcolumn}
\usepackage{hyperref} 
\hypersetup{colorlinks, allcolors=blue}

\begin{document}

\title{Exceeding the Chandrasekhar-Clogston limit in flat-band superconductors: A multiband strong-coupling approach}

\author{Kristian M{\ae}land} 
\affiliation{\mbox{Center for Quantum Spintronics, Department of Physics, Norwegian University of Science and Technology, NO-7491 Trondheim, Norway}} 
\author{Asle Sudb{\o}}
\email[Corresponding author: ]{asle.sudbo@ntnu.no}
\affiliation{\mbox{Center for Quantum Spintronics, Department of Physics, Norwegian University of Science and Technology, NO-7491 Trondheim, Norway}}

\begin{abstract}
Hybrid systems of superconductors and magnets display several intriguing properties, both from a fundamental physics point of view and with practical applications. Promising applications in superconducting spintronics motivate a search for systems where superconductivity can survive larger in-plane critical magnetic fields than the conventional limit. The Chandrasekhar-Clogston (CC) limit applies to thin-film conventional superconductors with in-plane magnetic fields such that orbital effects may be ignored. For a magnetic field strength comparable to the superconducting gap at zero temperature and zero field, a spin-split normal state attains lower free energy than the superconducting state. A multiband superconductor with a flat band placed just below the Fermi surface has been shown to surpass the CC limit using weak-coupling theory. Since the dimensionless coupling determining the critical temperature scales with the density of states, it is natural to anticipate corrections from strong-coupling theory in flat-band systems, owing to the large density of states of the flat bands. We derive Eliashberg equations and the free energy for a multiband superconductor in a magnetic field. First, we show that the CC limit can be exceeded by a small amount in one-band strong-coupling superconductors due to self-energy renormalization of the magnetic field. Next, we consider a two-band system with one flat band and find that the CC limit can be exceeded by a large amount also in strong-coupling theory, even when including hybridization between bands that intersect.    
\end{abstract}

\maketitle 

\section{Introduction}
The field of superconducting spintronics \cite{Linder2015Spintronic, Eschrig2011Spintronic, Eschrig2015SpintronicRev} aims to use proximity effects between magnets and superconductors (SCs) to realize dissipationless transport of information. Spin-split SCs are also of interest for the generation of spin-polarized supercurrents and the appearance of large thermoelectric effects, which convert excess heat into applicable energy \cite{Bergeret2018ThermoelectricRev, AliPRB2022, AliPRL2023}. Coexistence of superconductivity and magnetism is, however, a challenging prospect. 
Orbital effects \cite{Bergeret2018ThermoelectricRev, SFsuperconductivity, Yerin2013Dec, Wada1964ElishbergFreeEnergyT0} are suppressed in the case of in-plane magnetic fields in thin-film, effectively two-dimensional (2D), SCs. 
Nonetheless, at high enough magnetic field, a spin-split normal state has a lower free energy than the superconducting state, destabilizing the SC \cite{Chandrasekhar1962Sep, Clogston1962Sep, Sarma1963h>D, CClimitTriplet_Powell2003}. This is referred to as the Chandrasekhar-Clogston (CC) limit \cite{Chandrasekhar1962Sep, Clogston1962Sep} or the Pauli limit \cite{Salamone2023highB, Cao2021reentrantmoire}. The critical magnetic field at zero temperature is $h_c = \Delta_0/\sqrt{2} \approx 0.7 \Delta_0$, where $\Delta_0$ is the zero-temperature gap amplitude. 

When applying a magnetic field, the normal state gains energy by spin polarizing the bands around the Fermi surface (FS). The superconducting state, with a gap around the FS, cannot gain energy in the same way. The CC limit applies to conventional one-band SCs with Cooper pairs comprised of opposite-spin electrons \cite{CClimitTriplet_Powell2003}. The original papers by Chandrasekhar \cite{Chandrasekhar1962Sep} and Clogston \cite{Clogston1962Sep} considered spin-singlet superconductivity, while Ref.~\cite{CClimitTriplet_Powell2003} extends the CC limit to spin-triplet superconductivity with zero net spin Cooper pairs. On the other hand, spin-polarized spin triplet SCs are not affected by the Pauli pair breaking responsible for the CC limit \cite{CClimitTriplet_Powell2003}.

Finite-momentum Cooper pairing in Fulde-Ferrell-Larkin-Ovchinnikov (FFLO) states can surpass the CC limit \cite{FF, LO, CClimitTriplet_Powell2003, Islam2023Feb}. Additionally, multiband SCs with reentrant superconductivity \cite{Salamone2023highB, Cao2021reentrantmoire} and SCs driven out of equilibrium \cite{Ouassou2018HighBnonequi, Tjernshaugen2023Nov} can exceed the CC limit. Of particular interest in this paper, flat-band SCs also provide a way to exceed the CC limit for $s$-wave zero net spin pairing \cite{Ghanbari2022Feb}. Reference \cite{Ghanbari2022Feb} shows, using weak-coupling Bardeen-Cooper-Schrieffer (BCS) theory \cite{BCS}, that a flat band below the FS can boost the superconductivity in a dispersive band. This delays the transition to the normal state until the magnetic field strength corresponds to several times the zero-temperature gap. The flat band provides a larger condensation energy, while not contributing to any energy gain in the spin-split normal state until the magnetic field is large enough to move the flat band for one spin component above the FS \cite{Ghanbari2022Feb}. 
It is worth noting that while the coupling between the two crossing bands was considered, this analysis omitted the effects of band hybridization. In this paper, we incorporate band hybridization arising from self-energy effects, albeit in a simplified manner.

Flat bands can appear in partial line graphs \cite{FlatBandPartialLineGraphMiyahara2005}, e.g.~realized in octagraphene \cite{OctagrapheneSC_Li2022, Sheng2012Octagraphene, OctagrapheneCorrectBao2014, Nunes2020Jun}, in twisted bilayers such as twisted-bilayer graphene \cite{EliashbergFlatBandSchrodi2020Mar, EliashbergFlatBandSchrodi2021Apr, brekke2023interfacial, Andrei2020TBGrev, Marchenko2018ARPESflatBand, EPCTBGChoi2021}, in rhombohedral graphite \cite{EliashbergFlatBandOjajarvi2017, EliashbergFlatBandOjajarvi2018, Pierucci2015ARPESflatBand}, in Lieb lattices \cite{Lieb1989Mar, Brekke2023Aug, Okamoto2018Lieb, Slot2017Lieb}, and in diatomic kagome lattices \cite{Sethi2021May, Sethi2023May}. 
Often, the flat bands result from the interplay between several different hopping amplitudes in the nontrivial lattice structures. The electronic model we consider, with hybridization between a dispersive band and a flat band, bears similarities with the Anderson lattice model for Kondo insulators, only that we consider a doped version with the FS in the conduction band above the flat band \cite{Dzero2016KondoInsulatorRev, Lai2018KondoSemimetal}. In that sense, it is more similar to simple models for heavy-fermion SCs, though we consider phonon-driven superconductivity, while spin fluctuations are expected to dominate the superconducting pairing in heavy-fermion SCs \cite{Steglich2016Heavy-fermion}. A similar electronic model may also be realized in twisted-bilayer graphene \cite{Islam2023Feb}.

In the single-band case, the superfluid weight is proportional to the group velocity of the band \cite{PeottaTorma2015FlatBand}. Hence a single, completely flat band has zero superfluid weight. Since nonzero superfluid weight is essential for the Meissner effect and dissipationless flow, it represents a major caveat for flat-band SCs. However, in multiband systems, the superfluid weight has extra contributions related to quantum geometry, yielding a nonzero superfluid weight also for flat bands \cite{PeottaTorma2015FlatBand, Torma2022QuantumGeom}.

The BCS prediction for the critical temperature is $T_c \propto \exp(-1/\lambda)$, with dimensionless coupling $\lambda = V N_F$, where $V$ is the coupling constant and $N_F$ is the density of states (DOS) per spin on the FS. Weak-coupling theory applied to flat bands suggests that $T_c \propto \lambda$, eliminating the exponential suppression of $T_c$ in dispersive bands \cite{Miyahara2007FlatBandBCS, Kopnin2011FlatBandBCS, PeottaTorma2015FlatBand}. 
Nearly (completely) flat electronic bands give a very large (diverging) DOS. Naively this should mean stronger coupling, and hence, flat-band SCs offer a fruitful path toward high-$T_c$ superconductivity \cite{Aoki2020FlatBand}. Machine learning techniques applied to the search for candidate materials for high-$T_c$ superconductivity have yielded several materials with nearly flat bands just below the FS, even when the DOS was not an input \cite{Stanev2018MachineLearning}.

An increased coupling strength due to a large DOS also means that strong-coupling effects likely provide significant corrections to weak-coupling predictions in flat-band SCs \cite{EliashbergFlatBandOjajarvi2017}. Eliashberg \cite{Eliashberg1960Sep} introduced a strong-coupling theory of superconductivity, where electron self-energy effects and superconducting pairing are treated on the same footing \cite{EliashbergRevMarsiglio2020}. In that way, renormalizations of the electron bands caused by the same interaction which drives the superconducting pairing are included in the theoretical model. Often, Eliashberg theory is simplified by performing FS averages and neglecting vertex corrections \cite{EliashbergRevMarsiglio2020}. Both of these approximations rely on Migdal's theorem \cite{migdal1958interaction, Migdal2D}, namely that the electron bandwidth is much larger than the characteristic energy of the boson mediating the effective electron-electron interaction. Such a Migdal-Eliashberg theory is not applicable to flat bands, where the electron bandwidth is zero. References \cite{EliashbergFlatBandOjajarvi2017, EliashbergFlatBandOjajarvi2018, EliashbergFlatBandSchrodi2020Mar, EliashbergFlatBandSchrodi2021Apr} apply Eliashberg theory to systems with nearly flat bands by avoiding FS averages and instead considering simpler forms of the interaction.

The free energy of a normal metallic state, given information about its self-energy, was derived in Ref.~\cite{LuttingerWard1960Jun} using a variational approach. This was subsequently generalized to the superconducting state in Eliashberg theory \cite{Eliashberg1963, Bardeen1964ElishbergFreeEnergyT0, Wada1964ElishbergFreeEnergyT0, Carbotte1990FreeEnergy, Haslinger2003ChubukovFreeEnergy, Chubukov2022FEgeneral}.  
More recently, Ref.~\cite{Protter2021funkintFE} presented a way to derive the Eliashberg equations via functional integral methods. The derivation provides an expression for the free energy given in terms of solutions to the Eliashberg equations \cite{Lundemo2023, Aase2023Eliashberg}.
The latter approach is more easily generalizable to the case of multiple bands and the presence of a magnetic field and is therefore employed in this paper. Considerations of the free energy within Eliashberg theory are numerous \cite{Eliashberg1963, Bardeen1964ElishbergFreeEnergyT0, Wada1964ElishbergFreeEnergyT0, Carbotte1990FreeEnergy, Haslinger2003ChubukovFreeEnergy, Chubukov2023FENM, Chubukov2022FEgeneral, Chubukov2020FEspecialized, Yuzbashyan2022JulSpin, Yuzbashyan2022AugQCP, Yuzbashyan2022AugBreakdown, Protter2021funkintFE, Dalal2023Aug, Aase2023Eliashberg, Lundemo2023} but to the best of our knowledge no study of the CC limit within Eliashberg theory has been performed. 
One of the main results of this paper is an expression giving the free energy for a multiband system in a magnetic field in the spirit of Ref.~\cite{Protter2021funkintFE}.

References \cite{Linscheid2015magneticI, Linscheid2015magneticII, Aperis2015magneticField} study SCs in a magnetic field using Eliashberg theory, but do not discuss the free energy based on solutions to the Eliashberg equations. 
Therefore, after deriving the Eliashberg equations in Sec.~\ref{sec:Green} and presenting the free energy in Sec.~\ref{sec:FE}, we first consider a strong-coupling system with one dispersive band in Sec.~\ref{sec:1band}. We find that it is possible to exceed the CC limit slightly in a strong-coupling SC since self-energy effects effectively reduce the strength of the magnetic field. Section \ref{sec:2band} moves on to a two-band system as in Ref.~\cite{Ghanbari2022Feb}, and finds that the prediction of vastly exceeding the CC limit in flat-band SCs survives in strong-coupling Eliashberg theory. Section \ref{sec:2bandhybrid} shows that this applies also when including hybridization. We conclude in Sec.~\ref{sec:conclusion}, while the Appendixes offer additional details and the derivation of the free energy.

\section{Green's function derivation of Eliashberg equations} \label{sec:Green}

We are interested in the phase transition from a superconducting state to a normal state, driven by the Zeeman coupling to the electrons in an external magnetic field, in situations where orbital effects are suppressed. Typically, this is the case in thin-film superconductors with an in-plane magnetic field. Therefore, equilibrium considerations are sufficient, and the Matsubara formalism is a suitable choice in which to perform the many-body perturbation theory \cite{Eliashberg1961, EliashbergFlatBandOjajarvi2017, EliashbergFlatBandOjajarvi2018, ThingstadEliashberg, EliashbergRevMarsiglio2020, FetterWalecka, BruusFlensberg, abrikosov}. It is, however, applied to nonzero temperature, while the CC limit applies to zero temperature. Solving the Eliashberg equations at zero temperature is possible \cite{Eliashberg1960Sep, Schrieffer1963Apr, Schrieffer1964book, Scalapino1966Aug, Carbotte1990FreeEnergy}, but is considered more complicated than solving the nonzero-temperature Eliashberg equations. The complications arise from the appearance of singularities on the continuous real frequency axis which are challenging to handle correctly, while these are avoided on the discrete imaginary frequency axis \cite{Marsiglio1988AnalyticCont}. We therefore choose to approach the zero-temperature limit using the Matsubara formalism in this paper.

We consider the Hamiltonian
\begin{align}
\label{eq:Hamiltonian}
    H =& \sum_{\boldsymbol{k}l\sigma} \epsilon_{\boldsymbol{k}l\sigma} c_{\boldsymbol{k}l\sigma}^\dagger c_{\boldsymbol{k}l\sigma} + \sum_{\boldsymbol{q}} \omega_{\boldsymbol{q}}a_{\boldsymbol{q}}^\dagger a_{\boldsymbol{q}} \nonumber\\
    &+ \sum_{\boldsymbol{k}\boldsymbol{q}\sigma ll'} g_{\boldsymbol{q}} c_{\boldsymbol{k}+\boldsymbol{q},l',\sigma }^\dagger c_{\boldsymbol{k}l\sigma }(a_{\boldsymbol{q}}+a_{-\boldsymbol{q}}^\dagger),
\end{align}
describing a multiband electron system with a single phonon mode and band-independent electron-phonon coupling (EPC). We consider a 2D square lattice with $N$ lattice sites and periodic boundary conditions so that the (quasi)momentum $\boldsymbol{k}$ is restricted to a square first Brillouin zone (1BZ). With increasing $N$, this becomes a good model for the bulk. $\epsilon_{\boldsymbol{k}l\sigma} = \epsilon_{\boldsymbol{k}l}-\sigma h$ are the electron bands, $h$ is the external, in-plane magnetic field, and $c_{\boldsymbol{k}l\sigma}^\dagger$ creates an electron in band $l$ with momentum $\boldsymbol{k}$ and spin $\sigma$. The spins of the electrons are quantized along the external magnetic field, such that spin up points along the magnetic field. $\omega_{\boldsymbol{q}}$ is the phonon dispersion relation, while $a_{\boldsymbol{q}}^\dagger$ creates a phonon with momentum $\boldsymbol{q}$. $g_{\boldsymbol{q}}$ is the band-independent EPC strength, which is further assumed to only depend on the momentum transfer in the EPC. We employ units where $\hbar = k_B = 1$ and the lattice constant $a=1$.

In the Matsubara formalism, the Green's function for electrons is $G_{l}(\boldsymbol{k},\tau) = -\langle T_\tau \psi_{\boldsymbol{k}l}(\tau) \psi_{\boldsymbol{k}l}^\dagger (0) \rangle$, where $T_\tau$ is the time-ordering operator and $\psi_{\boldsymbol{k}l}^\dagger = (c_{\boldsymbol{k}l\uparrow}^\dagger, c_{\boldsymbol{k}l\downarrow}^\dagger, c_{-\boldsymbol{k}l\uparrow}, c_{-\boldsymbol{k}l\downarrow})$ is the Nambu spinor.
Here, $-\langle T_\tau \psi_{\boldsymbol{k}l}(\tau) \psi_{\boldsymbol{k}l}^\dagger (0) \rangle$ is a shorthand for $-\langle T_\tau \psi_{\boldsymbol{k}l}(\tau) \psi_{\boldsymbol{k}l}^\dagger (0) S(\beta, 0)\rangle_{\text{conn}}$, where 
\begin{equation}
    S(\tau, 0) = \sum_{n=0}^{\infty} \frac{(-1)^n}{n!} \int_0^{\tau} d\tau_1 \cdots d\tau_n T_\tau [H_{\text{int}}(\tau_1)\cdots H_{\text{int}}(\tau_n)].
\end{equation}
is the $S$ matrix and subscript ``conn'' means that only connected diagrams are counted \cite{BruusFlensberg, abrikosov, Schrieffer1964book}. $H_{\text{int}}(\tau)$ is the interaction Hamiltonian. In the latter definition of the Green's function, expectation values are calculated in the interaction picture \cite{BruusFlensberg}.

We set all expectation values involving operators from separate bands to zero. This is not strictly speaking true with the assumed form of EPC, and neglects band hybridization arising from, e.g., $-\langle T_\tau c_{\boldsymbol{k}l\uparrow}(\tau) c_{\boldsymbol{k}l'\uparrow}^\dagger (0) \rangle$ with $l\neq l'$. If the bands $\epsilon_{\boldsymbol{k}l\sigma}$ are isolated, the effects of hybridization terms are negligible. If two or more bands cross, neglecting hybridization could lead to erroneous predictions.

It is in principle possible to derive Eliashberg equations including hybridization effects from the self-energy, but the derivation would involve analytically inverting dense $2n\times 2n$ matrices with $n$ being the number of bands. Already for two bands this gives very involved analytic expressions. Additionally, $(4n)^2/2$ Eliashberg functions must be introduced to capture all effects, meaning that it will be very challenging to interpret the results. As a workaround, we instead consider prehybridized bare bands with simple assumptions on the hybridization in Sec.~\ref{sec:2bandhybrid}. In that case, hybridization effects can be ignored in the Eliashberg equations, since including them would lead to overcounting.

By setting interband expectation values to zero, we also ignore interband Cooper pairs, which are captured by, e.g., $-\langle T_\tau c_{-\boldsymbol{k},l,\uparrow}^\dagger(\tau) c_{\boldsymbol{k}l'\downarrow}^\dagger (0) \rangle$ with $l\neq l'$. An interband Cooper pair is formed of two electrons from separate bands. Interband Cooper pairs are usually ignored on the grounds that one cannot make zero momentum interband Cooper pairs in the case of disjoint FSs. 
Flat bands close to the FS offer states at all $\boldsymbol{k}$, so it is possible to construct zero-momentum interband Cooper pairs in our case. 
However, if there is hybridization between the dispersive band and the flat band, the part of the flat band that offers electrons with opposite momentum to those on the FS of the dispersive band moves further away from the FS. Hence the pairing of interband Cooper pairs becomes much weaker than intraband Cooper pairs. We therefore ignore interband Cooper pairs in this paper.   
Reference \cite{Salamone2023highB} considers a case where the formation of interband Cooper pairs leads to reentrant superconductivity above the CC limit in a two-band system, a situation that could also be relevant in our system. In this paper, we do not focus on reentrant superconductivity due to interband Cooper pairs. Rather, we study how large the magnetic field needs to be to break intraband Cooper pairs.

The Fourier transform (FT) from imaginary time to imaginary frequency is
\begin{equation}
\label{eq:GFT}
    G(\boldsymbol{k}, i\omega_n)  = \int_0^\beta d\tau e^{i\omega_n\tau} G(\boldsymbol{k},\tau),
\end{equation}
with inverse temperature $\beta = 1/T$ and Matsubara frequencies $i\omega_n = i(2n+1) \pi T$ for fermions and $i\omega_\nu = i2\nu \pi T$ for bosons.
Ignoring interactions yields the  bare electron Green's function,
\begin{equation}
    G_{0l}(k) = (i\omega_n \tau_0 \sigma_0 -\epsilon_{\boldsymbol{k}l} \tau_3 \sigma_0 + h \tau_3 \sigma_3)^{-1},
\end{equation}
where $k = (\boldsymbol{k},i\omega_n)$, $\tau_0$ and $\sigma_0$ are unit matrices, and $\tau_i$ and $\sigma_i$ for $i=\{1,2,3\}$ are Pauli matrices. $\tau_i$ applies to the particle-hole degree of freedom, while $\sigma_i$ covers the spin degree of freedom. $\tau_i \sigma_j$ is a shorthand for the outer product $\tau_i \otimes \sigma_j$, meaning that they are $4\times4$ matrices.

\begin{figure}
    \centering
    \includegraphics[width = 0.9\linewidth]{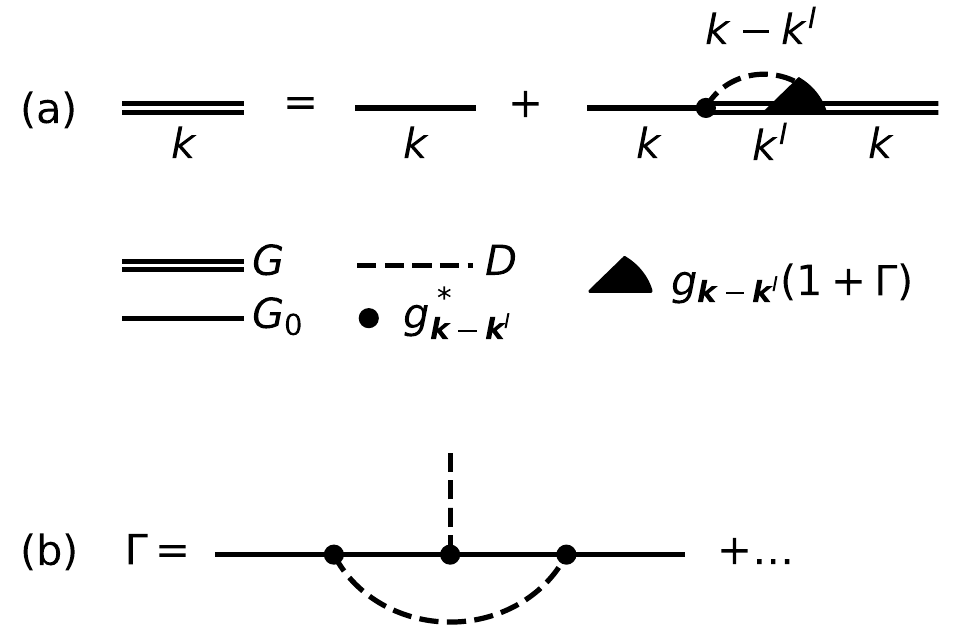}
    \caption{(a) Feynman diagrams illustrate the renormalized Green's function $G = G_0 + G_0 \Sigma G$, defined self consistently in terms of the bare electron Green's function $G_0$ and the phonon driven electron self-energy $\Sigma$. Band indices are suppressed, while $k = \boldsymbol{k}, i\omega_n$. Solid lines are electron Green's functions, and dashed lines are phonon propagators $D$. A vertex correction $\Gamma$ is included in (a). In (b), we show the lowest order vertex correction. The external legs are not a part of $\Gamma$ but are included for clarity.}
    \label{fig:Feynman}
\end{figure}

EPC affects the electron Green's function $G_l(k)$ through the self-energy $\Sigma(k)$. Generally, the self-energy depends on the band index of the external bands. However, under the assumption that EPC is band independent, the self-energy is independent of the external band indices. The Dyson equation gives $G_l^{-1}(k) = G_{0l}^{-1}(k)-\Sigma(k)$. From an $S$-matrix expansion \cite{BruusFlensberg, ThingstadEliashberg, Maeland2021Sep}, the self-energy is 
\begin{equation}
\label{eq:EPCselfenergy}
    \Sigma(k) = -\sum_{k'l} |g_{\boldsymbol{k}-\boldsymbol{k}'}|^2 D(k-k') \tau_3 \sigma_0 G_l(k') \tau_3 \sigma_0,
\end{equation}
where $\sum_{k} = T \sum_{\boldsymbol{k},i\omega_{n}}$. Figure \ref{fig:Feynman}(a) illustrates the Dyson equation on the form $G = G_0 + G_0 \Sigma G$.
Note that Eq.~\eqref{eq:EPCselfenergy} is a self-consistent definition of the self-energy, where $D(q) = D(\boldsymbol{q},i\omega_\nu)$ and $G_l(k)$ are themselves renormalized Green's functions for the phonon and electron, respectively. Nevertheless, certain vertex corrections have been ignored, as illustrated in Fig.~\ref{fig:Feynman}(b). 
Reference \cite{EliashbergFlatBandSchrodi2021Apr} shows that the vertex correction in Fig.~\ref{fig:Feynman}(b) is negligible in flat-band systems, given that the flat band is much closer to the FS than the characteristic phonon energy. 

In this paper, we use a simplified version of the renormalized phonon Green's function, namely the bare Green's function for Einstein phonons $\omega_{\boldsymbol{q}} = \omega_E$,
\begin{equation}
    D(q) = D(i\omega_\nu) = \frac{-2\omega_E}{\omega_\nu^2+\omega_E^2}.
\end{equation}
We expect that our results become more accurate the closer the actual $D(q)$ is to this approximation. Appealing to universality, the results should give qualitative predictions for all EPC-driven SCs as the main point is that there exists a phonon mediated attractive interaction between electrons. Furthermore, we are ignoring the phonon self-energy because in a real system, the renormalized phonon spectrum can be measured by spectroscopic methods, and used as an input in theoretical calculations. Including phonon self-energy effects would then amount to overcounting \cite{EliashbergFlatBandOjajarvi2017}. 
Viewing the phonons as already renormalized also means that the dimensionless coupling strength $\lambda$ is renormalized. The renormalized $\lambda$ has a higher upper limit where Eliashberg theory is expected to break down, than the bare $\lambda_0$ \cite{Yuzbashyan2022AugBreakdown, Chubukov2020FEspecialized, Chubukov2022FEgeneral}.

We consider EPC-driven superconductivity, and do not explicitly include Coulomb repulsion. When included in Eliashberg theory, the effect of the Coulomb interaction is essentially just to reduce the interaction strength slightly \cite{Schrieffer1964book, EliashbergRevMarsiglio2020}. The calculations in Ref.~\cite{MorelAnderson1962} show that the Coulomb repulsion can be treated as a small pseudopotential \cite{EliashbergRevMarsiglio2020}. This is due to the retarded nature of the EPC compared with the direct nature of the Coulomb repulsion \cite{EliashbergRevMarsiglio2020}.

Since EPC conserves spin, we assume that only the spin conserving expectation values are nonzero in $G_{l}(\boldsymbol{k},\tau) = -\langle T_\tau \psi_{\boldsymbol{k}l}(\tau) \psi_{\boldsymbol{k}l}^\dagger (0) \rangle$, such that only half the matrix elements are nonzero. This ignores equal spin Cooper pairs, which are not expected to appear in EPC-driven SCs. From the Dyson equation $G_l^{-1}(k) = G_{0l}^{-1}(k)-\Sigma(k)$, $\Sigma(k)$ must take the same form as $G_l(k)$. This limits the description to eight outer products $\tau_i \sigma_j$. We write
\begin{widetext}
\begin{equation}
\label{eq:selfenergyparam}
    \Sigma = (1-Z)i\omega_n \tau_0 \sigma_0 + \eta \tau_0 \sigma_3 + \chi \tau_3 \sigma_0 - \Sigma_h \tau_3 \sigma_3 + \phi_e^R \tau_2\sigma_2 + \phi_e^I \tau_1 \sigma_2 + \phi_o^R \tau_1 \sigma_1 + \phi_o^I \tau_2 \sigma_1,
\end{equation}
Here, the $k$ dependence of the Eliashberg functions, $Z = Z(k)$ etc., is suppressed in the notation. This parametrization yields
\begin{equation}
    G_l^{-1} = G_{0l}^{-1} - \Sigma = i\omega_n Z \tau_0 \sigma_0 -(\epsilon_{\boldsymbol{k}l}+\chi)\tau_3 \sigma_0  - \eta \tau_0 \sigma_3 +(h+\Sigma_h)\tau_3\sigma_3 -  \phi_e^R \tau_2\sigma_2 - \phi_e^I \tau_1 \sigma_2 - \phi_o^R \tau_1 \sigma_1 - \phi_o^I \tau_2 \sigma_1.
\end{equation}
To invert $G_l^{-1}(k)$, we identify two $2\cross2$ blocks, with determinants
\begin{equation}
    \Theta_{1l}(k) = [i\omega_n Z(k) + h + \Sigma_h(k)]^2 - [\epsilon_{\boldsymbol{k}l}+\chi(k) +\eta(k)]^2-[ \phi_e^R(k)- \phi_o^R(k)]^2 - [ \phi_e^I(k) +  \phi_o^I(k)]^2,
\end{equation}
\begin{equation}
    \Theta_{2l}(k) = [i\omega_n Z(k) - h - \Sigma_h(k)]^2 - [\epsilon_{\boldsymbol{k}l}+\chi(k) -\eta(k)]^2-[ \phi_e^R(k)+ \phi_o^R(k)]^2 - [ \phi_e^I(k) -  \phi_o^I(k)]^2,
\end{equation}
for the outer block and the inner block, respectively. The Green's function is then
\begin{equation}
\label{eq:Green}
    G_l = \begin{pmatrix}
    \frac{i\omega_n Z +\epsilon_{\boldsymbol{k}l}+\chi +\eta + h+ \Sigma_h}{\Theta_{1l}} & 0 & 0 & \frac{-\phi_e^R-i\phi_e^I +\phi_o^R-i\phi_o^I}{\Theta_{1l}} \\ 
    0 & \frac{i\omega_n Z +\epsilon_{\boldsymbol{k}l}+\chi -\eta - h- \Sigma_h}{\Theta_{2l}}   & \frac{\phi_e^R+i\phi_e^I +\phi_o^R-i\phi_o^I}{\Theta_{2l}} & 0 \\ 
    0 & \frac{\phi_e^R-i\phi_e^I +\phi_o^R+i\phi_o^I}{\Theta_{2l}} & \frac{i\omega_n Z -\epsilon_{\boldsymbol{k}l}-\chi +\eta - h- \Sigma_h}{\Theta_{2l}}  & 0 \\  
    \frac{-\phi_e^R+i\phi_e^I +\phi_o^R+i\phi_o^I}{\Theta_{1l}} & 0 & 0 &  \frac{i\omega_n Z -\epsilon_{\boldsymbol{k}l}-\chi -\eta + h+ \Sigma_h}{\Theta_{1l}}
    \end{pmatrix}.
\end{equation}
The poles in the Green's function, $\Theta_{il}(k) = 0$, give information about renormalized bands. The poles are
\begin{equation}
    i\omega_n = \frac{-h-\Sigma_h(k)}{Z(k)} \pm \bqty{\pqty{\frac{\epsilon_{\boldsymbol{k}l}+\chi(k) +\eta(k)}{Z(k)}}^2 + \pqty{\frac{\phi_e^R(k)- \phi_o^R(k)}{Z(k)}}^2 + \pqty{\frac{\phi_e^I(k) +  \phi_o^I(k)}{Z(k)}}^2}^{1/2},
\end{equation}
\begin{equation}
    i\omega_n = \frac{h+\Sigma_h(k)}{Z(k)} \pm \bqty{\pqty{\frac{\epsilon_{\boldsymbol{k}l}+\chi(k) -\eta(k)}{Z(k)}}^2 + \pqty{\frac{\phi_e^R(k)+ \phi_o^R(k)}{Z(k)}}^2 + \pqty{\frac{\phi_e^I(k) -  \phi_o^I(k)}{Z(k)}}^2}^{1/2}.
\end{equation}
Comparing to the BCS spectrum $E_{\boldsymbol{k}l\sigma} = -\sigma h \pm \sqrt{\epsilon_{\boldsymbol{k}l}^2 + \Delta^2}$ \cite{Ghanbari2022Feb}, we can define the superconducting gaps as the first complex roots of
\begin{equation}
    [\Delta_1(k)]^2  = \pqty{\frac{\phi_e^R(k)- \phi_o^R(k)}{Z(k)}}^2 + \pqty{\frac{\phi_e^I(k) +  \phi_o^I(k)}{Z(k)}}^2, \quad [\Delta_2(k)]^2 = \pqty{\frac{\phi_e^R(k)+ \phi_o^R(k)}{Z(k)}}^2 + \pqty{\frac{\phi_e^I(k) -  \phi_o^I(k)}{Z(k)}}^2.
\end{equation}
\end{widetext}
However, self-energies given on the imaginary frequency axis are not as easy to interpret or connect with experiments. The quasiparticle energy is defined in terms of the real-axis frequency results \cite{Marsiglio1988AnalyticCont}.
Technically, these equations should be analytically continued \cite{Marsiglio1988AnalyticCont, EliashbergFlatBandSchrodi2020Mar, EliashbergFlatBandSchrodi2021Apr} to real frequencies, after which renormalized bands can be found by solving the equations self-consistently. 
We leave the renormalized bands at this approximate level and mostly consider the zero-imaginary-frequency limit of these functions when discussing renormalization effects.
By the zero-imaginary-frequency limit, we mean the average value of these functions at $i\omega_{n=-1} = -i\pi T$ and $i\omega_{n=0} = i \pi T$, which, when $T\to 0$ should approach the zero real frequency limit.
Continuing the comparison with the BCS spectrum,
\begin{equation}
\label{eq:heff}
    h_{\text{eff}}(k) = \frac{h+\Sigma_h(k)}{Z(k)}
\end{equation}
represents a renormalized magnetic field due to self-energy effects, while $Z(k)$ is a mass renormalization. $\chi(k)$ and $\eta(k)$ renormalize the electron bands, where $\eta(k)$ represents the spin-dependent part. 

Inserting the Green's function in Eq.~\eqref{eq:Green} into the EPC self-energy in Eq.~\eqref{eq:EPCselfenergy} and comparing with the parametrization of the self-energy in Eq.~\eqref{eq:selfenergyparam}, yields the Eliashberg equations given in Appendix \ref{app:fullEliashberg}. 
By considering symmetries in the expectation values involved in $G_{l}(\boldsymbol{k},\tau) = -\langle T_\tau \psi_{\boldsymbol{k}l}(\tau) \psi_{\boldsymbol{k}l}^\dagger (0) \rangle$ a set of symmetries are found for the matrix elements. Translated to the Eliashberg functions, these are $f(\boldsymbol{k}, i\omega_n) = f(\boldsymbol{k}, -i\omega_n)^*$ for all the Eliashberg functions under inversion of frequency. When inverting both momentum and frequency, $Z(k) = Z(-k)$, $\chi(k) = \chi(-k)$, $\Sigma_h(k) = \Sigma_h(-k)$, $\phi_e^R(k) = \phi_e^R(-k)$, $\phi_e^I(k) = \phi_e^I(-k)$, $\eta(k) = -\eta(-k)$, $\phi_o^R(k) = -\phi_o^R(-k)$, and $\phi_o^I(k) = -\phi_o^I(-k)$. 

For the special cases in which the Eliashberg functions are even in momentum, or momentum independent, the combination of the above symmetries leads to $Z, \chi, \Sigma_h, \phi_e^{R/I} \in \mathbb{R}$, while $\eta, \phi_o^{R/I}$ are purely imaginary. As a result, $\phi_e^R + i \phi_e^I$ is a complex, even momentum, spin singlet, even frequency superconducting gap. Meanwhile, $\phi_o^R + i \phi_o^I$ is a complex, even momentum, spin triplet, odd frequency superconducting gap \cite{Linder2019Oddw}. The spin symmetry of the gaps is gleaned from the elements of the Green's function matrix in Eq.~\eqref{eq:Green} and its definition $G_{l}(\boldsymbol{k},\tau) = -\langle T_\tau \psi_{\boldsymbol{k}l}(\tau) \psi_{\boldsymbol{k}l}^\dagger (0) \rangle$. The Eliashberg equations for $\phi_e^I$ and $\phi_o^I$ are the same as those for $\phi_e^R$ and $\phi_o^R$. This is because both the even- and odd-frequency superconducting gaps can be multiplied by the same global phase factor $e^{i\theta}$. We choose the phase such that $\phi_e^I = \phi_o^I = 0$ and rename $\phi_e^R = \phi_e$ and $\phi_o^R = \phi_o$. Then, $\Delta_2 = \Delta_1^*$ so for simplicity we discuss $\Delta(i\omega_n) \equiv \Delta_1(i\omega_n)$ for the remainder of the paper.

Eliashberg equations are typically simplified using FS averages to integrate out the momentum degree of freedom leaving a factor of the DOS on the FS. Since we intend to consider flat-band systems, this is not expected to give good approximations. Flat bands give large peaks in the DOS so that the DOS on the FS is not representative. Therefore, we take an alternative route to simplify the Eliashberg equations. We assume that the phonons are dispersionless Einstein phonons $\omega_{\boldsymbol{q}} = \omega_E$ and that the coupling $g_{\boldsymbol{q}}$ is isotropic, $g_{\boldsymbol{q}} = g/\sqrt{N}$. The same model was used for flat-band systems in Refs.~\cite{EliashbergFlatBandSchrodi2020Mar, EliashbergFlatBandSchrodi2021Apr} and similar models have been applied in numerous systems \cite{Chubukov2022FEgeneral, Dalal2023Aug, Yuzbashyan2022JulSpin, Schrodi2021IsotropicEPC, EliashbergFlatBandOjajarvi2017, EliashbergFlatBandOjajarvi2018}. 
Simplifying to Einstein phonons with isotropic coupling removes the $\boldsymbol{k}$ dependence on the right-hand side of the Eliashberg equations in Appendix \ref{app:fullEliashberg}, and then the Eliashberg functions depend only on frequency. This gives the Eliashberg equations 
\begingroup
\allowdisplaybreaks
\begin{align}
    Z(i\omega_n) =& 1+\frac{g^2}{2N i\omega_{n}}\sum_{k'l}  D_{nn'}\nonumber\\
    &\times \bqty{\frac{i\omega_{n'} Z(i\omega_{n'})}{\Theta_{l}^{+}(k')} + \frac{h + \Sigma_h(i\omega_{n'})}{\Theta_{l}^{-}(k')}},  \\
    \Sigma_h(i\omega_n) =& \frac{g^2}{2N}\sum_{k'l} D_{nn'}\bqty{\frac{i\omega_{n'} Z(i\omega_{n'})}{\Theta_{l}^{-}(k')} + \frac{h + \Sigma_h(i\omega_{n'})}{\Theta_{l}^{+}(k')}}, \\
    \chi(i\omega_n) =& \frac{-g^2}{2N}\sum_{k'l} D_{nn'}\bqty{\frac{\epsilon_{\boldsymbol{k}'l}+\chi(i\omega_{n'})}{\Theta_{l}^{+}(k')} +\frac{\eta(i\omega_{n'})}{\Theta_{l}^{-}(k')} }, \\
    \eta(i\omega_n) =& \frac{-g^2}{2N}\sum_{k'l} D_{nn'}\bqty{\frac{\epsilon_{\boldsymbol{k}'l}+\chi(i\omega_{n'})}{\Theta_{l}^{-}(k')} +\frac{\eta(i\omega_{n'})}{\Theta_{l}^{+}(k')} }, \\
    \phi_e(i\omega_n) =& \frac{g^2}{2N}\sum_{k'l} D_{nn'}\bqty{\frac{\phi_e(i\omega_{n'})}{\Theta_{l}^{+}(k')} - \frac{\phi_o(i\omega_{n'})}{\Theta_{l}^{-}(k')}}, \\
    \phi_o(i\omega_n) =& \frac{g^2}{2N}\sum_{k'l} D_{nn'}\bqty{-\frac{\phi_e(i\omega_{n'})}{\Theta_{l}^{-}(k')} + \frac{\phi_o(i\omega_{n'})}{\Theta_{l}^{+}(k')}},
\end{align}
\endgroup
with $D_{nn'} = D(i\omega_n-i\omega_{n'})$ and $1/\Theta_{l}^\pm (k) = 1/\Theta_{1l}(k) \pm 1/\Theta_{2l}(k)$. 

Momentum-independent Eliashberg functions present a significant simplification. Still, the $\boldsymbol{k}'$ sum on the right-hand side cannot be integrated out for flat-band systems. 
Instead, we compute the sum directly, capturing the full electron bandwidth. Hence, this goes beyond the standard Migdal-Eliashberg theory \cite{EliashbergFlatBandSchrodi2020Mar, EliashbergFlatBandSchrodi2021Apr, EliashbergRevMarsiglio2020}. We increase $N$, the number of evenly spaced $\boldsymbol{k}$ values in the 1BZ, until results become essentially independent of $N$, limiting finite-size effects. As a result, solving the above equations is more computationally demanding than solving the standard Migdal-Eliashberg equations. 
We compute the frequency sum on the right-hand side as a convolution using  the fast Fourier transformation. The sum has an infinite number of terms, but contributions from large $|\omega_{n'}|$ should be small. We introduce a cutoff $|\omega_n| <M$ and check that $M$ is large enough so that the results do not depend on $M$. 
The self-consistent Eliashberg equations are solved by fixed-point iteration. We place a tolerance on the relative difference between successive iterations to judge when the result converges.

\section{Free energy} \label{sec:FE}
Appendix \ref{app:funkint} derives the free energy using functional integral methods. The expression for the free energy is
\begin{align}
    F =& -\sum_{k l} \ln[ \beta^2 \Theta_{1l}(k) ] +\sum_{kk' ll'}\lambda(i\omega_{n}-i\omega_{n'})  \nonumber\\ 
    &\times\bigg[ \frac{\Bar{\phi}(i\omega_n)\phi(i\omega_{n'})}{\Theta_{1l}(k)\Theta_{1l'}(k')}-\frac{1}{2}\frac{G_{l\downarrow}^{-1}(-k)G_{l'\downarrow}^{-1}(-k')}{\Theta_{1l}(k)\Theta_{1l'}(k')}\nonumber \\
    &-\frac{1}{2}\frac{G_{l\uparrow}^{-1}(-k)G_{l'\uparrow}^{-1}(-k')}{\Theta_{2l}(k)\Theta_{2l'}(k')}\bigg]. \label{eq:FE}
\end{align}
Here, $\lambda(i\omega_\nu) = -g^2 D(i\omega_\nu)/N$, $G_{l\sigma}^{-1}(k) = G_{0l\sigma}^{-1}(k)-i\Sigma^\sigma(i\omega_n)$, $G_{0l\sigma}^{-1}(k) = i\omega_n - \epsilon_{\boldsymbol{k}l}+\sigma h$,
\begin{align*}
    i\Sigma^\uparrow (i\omega_n) &= [1- Z(i\omega_n)]i\omega_n +\chi(i\omega_n)+\eta(i\omega_n)-\Sigma_h(i\omega_n), \\
    i\Sigma^\downarrow (i\omega_n) &= [1-Z(i\omega_n)]i\omega_n +\chi(i\omega_n)-\eta(i\omega_n)+\Sigma_h(i\omega_n), \\
    \phi(i\omega_n) &= \bar{\phi}(i\omega_n) = \phi_e(i\omega_n)-\phi_o(i\omega_n).
\end{align*}
The symmetries of the Eliashberg functions under inversion of $i\omega_n$ ensure that the free energy is a real quantity.
As explained in Appendix \ref{app:funkint}, the expression for the free energy has an arbitrary zero point, so that only free energy differences are meaningful. We define $\Delta F = F_s -F_n$ as the free energy difference between the superconducting state and the normal state. $\Delta F < 0$ indicates that the superconducting state is energetically preferred, while the normal state is preferred if $\Delta F > 0$. The normal state is a solution of the Eliashberg equations where $\phi_e = \phi_o = 0$, and the superconducting state has either $\phi_e$ or $\phi_o$ nonzero, or both nonzero. 

Ideally, we should show that the superconducting state is the global minimum of the free energy. However, using the functional integral formalism the free energy is derived from a complex action, where the concept of a global minimum has no meaning. Reference \cite{Dalal2023Aug} suggests searching for saddle points in the complex action instead of a global minimum of the free energy. Meanwhile, Ref.~\cite{Chubukov2022FEgeneral} suggests imposing symmetries on the Eliashberg functions also away from stationary points, such that the action remains real and can be viewed as a free energy. We consider FSs with no nesting vectors so that spin-density wave correlations will not be a competing order \cite{Fjaerbu2019arneAFMNM, ThingstadEliashberg}. We therefore assume that the normal and superconducting states are the only possible states, in which case the sign of $\Delta F$ determines the global minimum of the free energy.

\section{One dispersive band} \label{sec:1band}
\begin{figure}
    \centering
    \includegraphics[width = \linewidth]{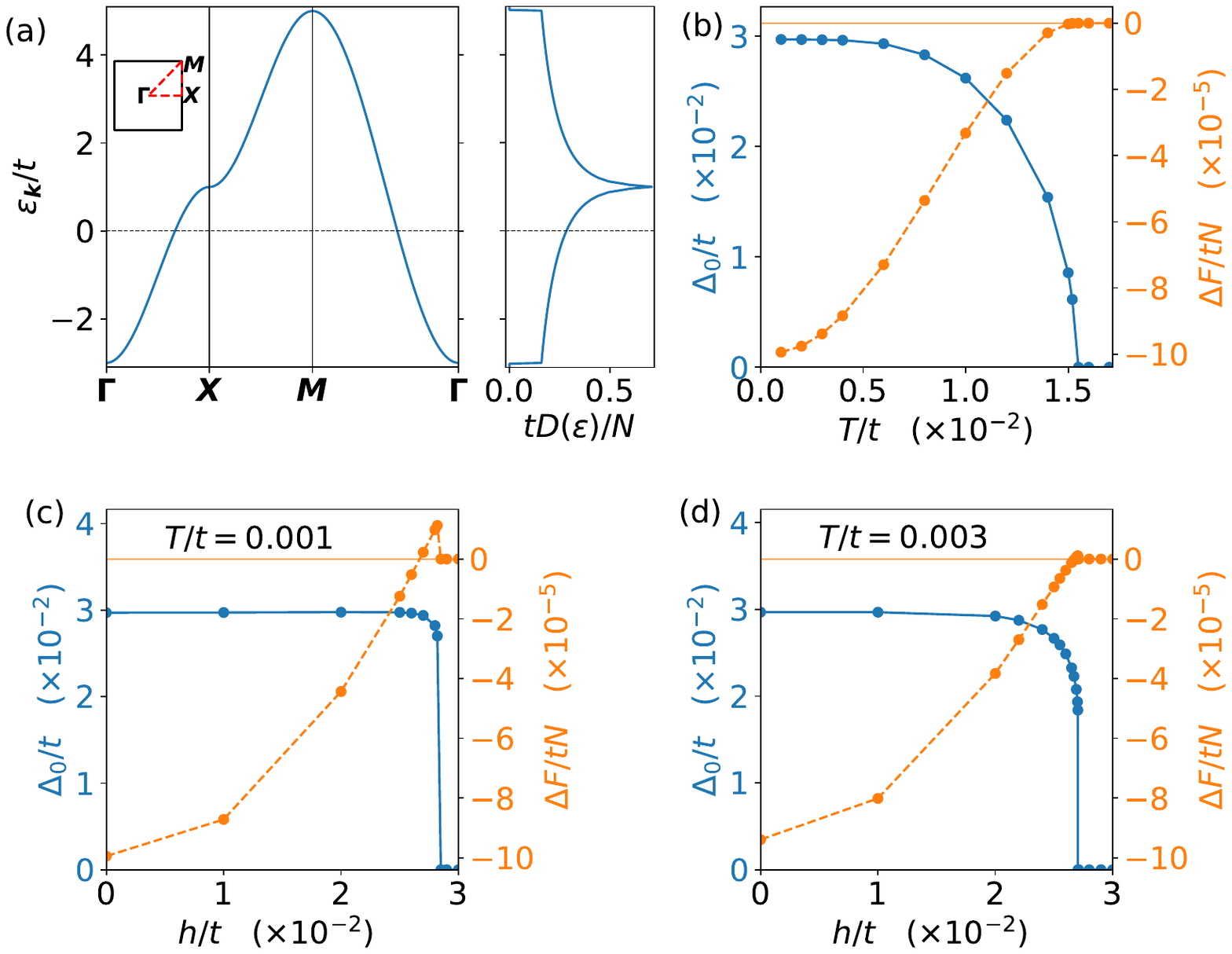}
    \caption{(a) The electron dispersion for a tight-binding model on the square lattice along the path sketched in the inset for $h=0$. The right panel shows the density of states. (b) The superconducting gap $\Delta_0 = \Re\Delta(i\omega_{n=0})$ (solid blue curve) as a function of temperature at $h=0$, giving $T_c/t \approx 0.015$ and $\Delta_0/t \approx 0.030$. The free energy difference between the superconducting and normal state $\Delta F$ is shown as a dashed orange curve. Solid circles indicate the calculated points. (c) $\Delta_0$ as a function of magnetic field (solid blue curve), and $\Delta F$ (dashed orange curve) for $T/t = 0.001$. A nonzero gap is shown as long as a nonzero solution is found to the Eliashberg equations. The sign of $\Delta F$ determines which state is preferred. (d) Same as (c), but for $T/t = 0.003$. In both cases, $\Delta F$ becomes positive for $h_c/t \approx 0.027$ giving $h_c/\Delta_0 \approx 0.9$. When $\Delta_0$ drops to zero as a function of temperature or magnetic field the superconducting state is the same as the normal state, and so $\Delta F = 0$. Parameters are $\mu = -t$, $g/t = 0.7$, $\omega_E/t = 0.2$, $M = 20\omega_E$, $N = 200^2$, and tolerance for convergence $10^{-5}$. }
    \label{fig:DF1}
\end{figure}

Since the CC limit to our knowledge has not been investigated using Eliashberg theory before, we first consider one dispersive band. We take the tight binding dispersion on the square lattice, $\epsilon_{\boldsymbol{k}} = -\mu-2t(\cos k_x + \cos k_y)$ with $\mu$ being the chemical potential and $t$ being the hopping parameter \cite{ThingstadEliashberg}. Electrons have spin, so what we refer to as one band is actually two spin degenerate bands in zero magnetic field. In a nonzero magnetic field, we get two spin split bands $\epsilon_{\boldsymbol{k}\sigma} = \epsilon_{\boldsymbol{k}}-\sigma h$. Figure \ref{fig:DF1}(a) shows the dispersive band and the DOS $D(\epsilon) = \sum_{\boldsymbol{k}\sigma}\delta(\epsilon-\epsilon_{\boldsymbol{k}\sigma})$ at zero magnetic field.

\begin{figure}
    \centering
    \includegraphics[width = \linewidth]{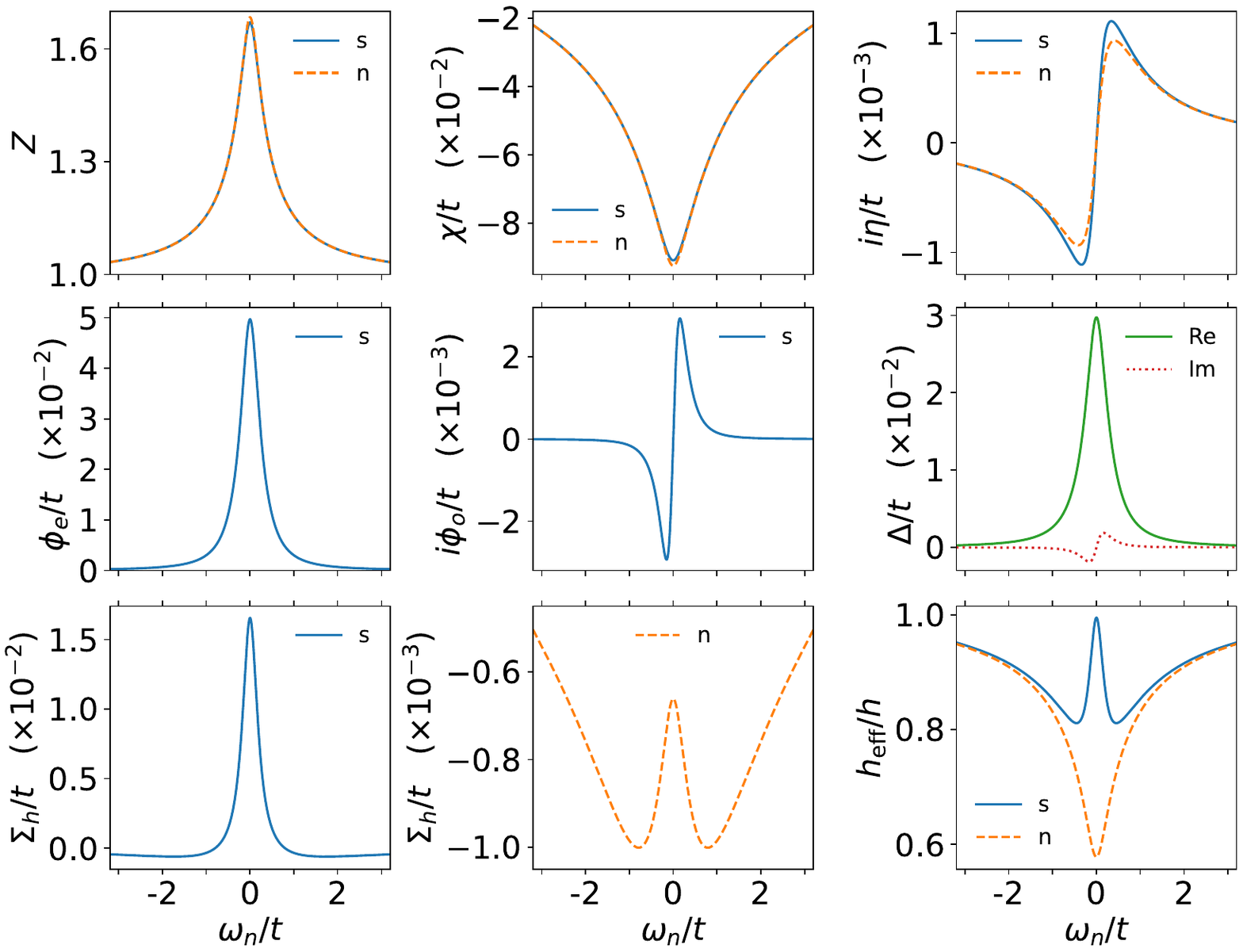}
    \caption{Solutions to the Eliashberg equations for the superconducting state (s; solid blue curves) and normal state (n; dashed orange curves) with one dispersive band. The superconducting gap $\Delta(i\omega_n)$ is shown with the real part represented by a solid green curve and the imaginary part represented by a dotted red curve. The gap is zero in the normal state. We also show the effective magnetic field $h_{\text{eff}}(i\omega_n) = [h+\Sigma_h(i\omega_n)]/[Z(i\omega_n)]$ in the superconducting and normal states. Parameters are $h/t = 0.025$, $T/t = 0.001$, and otherwise the same as in Fig.~\ref{fig:DF1}. }
    \label{fig:E1}
\end{figure}

Since the weak-coupling limit of Eliashberg theory reproduces BCS predictions \cite{EliashbergFlatBandOjajarvi2017, Chubukov2020FEspecialized}, we suspect the same holds for the CC limit. Therefore, we consider a strong EPC to see if the effects of renormalization change the CC limit for strong-coupling SCs. Figure \ref{fig:DF1} shows the superconducting gap and the free energy difference as a function of temperature and magnetic field. We find $T_c/t \approx 0.015$, $\Delta_0/t \approx 0.030$, and $h_c/t \approx 0.027$ for both $T/T_c \approx 1/5$ and $T/T_c \approx 1/15$. This yields $h_c/\Delta_0 \approx 0.9$, which is larger than the CC limit. With the choice of parameters in Fig.~\ref{fig:DF1}, we are in a range where vertex corrections, such as the one in Fig.~\ref{fig:Feynman}(b), are negligible, since $\omega_E \ll 8t$ \cite{migdal1958interaction, Migdal2D}. The results in Fig.~\ref{fig:DF1} show that the temperature dependence of both the gap amplitude $\Delta_0$ and the critical magnetic field $h_c$ converges for $T<T_c/5$. This indicates that taking the low-temperature limit within the Matsubara formalism gives good approximations to the zero temperature case. 

Note that $h_c$ is where $\Delta F$ changes sign, not where nonzero solutions of the gap cease to exist. At low temperatures, the phase transition from the superconducting state to the normal state is first order since $\Delta F$ changes sign while $\Delta_0 (h)$ is nonzero. Hence, the gap drops discontinuously to zero at $h_c$. At higher temperatures, we find that nonzero solutions to the gap cease to exist when $\Delta F$ is still negative indicating a second order, continuous phase transition. The same result was found in Ref.~\cite{Aperis2015magneticField} without considering the free energy. Hence our results offer an alternative perspective on the low-temperature first-order phase transition. It is caused by a sign change in the free energy difference, not a sudden stop in obtainable solutions of the Eliashberg equations with a nonzero gap. This is the same situation found using BCS theory \cite{Ghanbari2022Feb}.

Figure \ref{fig:E1} shows solutions to the Eliashberg equations. 
We note strong renormalization effects in the strong-coupling case, as expected. The magnetic field renormalization $\Sigma_h$, the spin-dependent renormalization $\eta$, and the odd frequency part of the gap $\phi_o$ are only nonzero in the presence of a magnetic field. Like Ref.~\cite{Aperis2015magneticField}, we find magnetic-field-induced odd-frequency superconductivity that coexists with a dominating even-frequency correlation. 
Since the even-frequency gap is dominating, we expect that the odd-frequency gap has small effects on the critical magnetic field.

As shown in Fig.~\ref{fig:E1}, $h_{\text{eff}}(i\omega_n) < h$, i.e., renormalizations reduce the effective magnetic field in the SC, yielding a lower band splitting than the external magnetic field would yield in BCS theory. This permits $h_c/\Delta_0 > 0.7$ in strong-coupling SCs. However, the CC limit is not exceeded by a large amount. Note that the effective magnetic field is in general different in the normal state and the superconducting state. It is natural to assume that the actual band splitting in the SC still follows the CC limit, but the external magnetic field is the relevant parameter that is tuned in an experiment. The band splitting, though measurable in angle-resolved photoemission spectroscopy (ARPES) \cite{Maeland2021Sep}, is less tunable. Since $h_{\text{eff}}(i\omega_n)$ depends on the parameters in the system, the CC limit for strong-coupling SCs becomes material dependent, as opposed to the general CC limit for weak-coupling SCs. It is worth noting that in certain applications, such as thermoelectric effects \cite{Bergeret2018ThermoelectricRev, AliPRB2022, AliPRL2023}, a large band splitting in an SC is the desired situation, not necessarily a larger magnetic field. In all the systems we considered, $h_{\text{eff}}(i\omega_n) < h$, but it seems in principle possible that $h_{\text{eff}}(i\omega_n) > h$ due to self-energy effects given that $\Sigma_h$ is positive and large enough.

We have defined $\Delta_0 \approx \Re\Delta(i\omega_{n=0} = i\pi T)$, where $\Delta_0$ is the zero-temperature gap amplitude. As mentioned, the results should technically be analytically continued to the real axis to compare with experiments. We therefore discuss whether $\Delta_0 \approx \Re\Delta(i\omega_{n=0})$ is a reasonable approximation. As $T\to 0$, $i\omega_{n=0} \to 0$ and so $\Delta(i\omega_{n=0}) \to \Delta(0)$. 
We find that $\Re\Delta(i\omega_n)$ is relatively smooth around zero frequency with a negative second derivative. Also, its zero-imaginary-frequency limit converges when decreasing $T$. 
Meanwhile, $\Im\Delta(i\omega_n)$ is odd so that its zero-frequency limit is zero. Therefore $\Re\Delta(i\omega_{n=0})$ at a low temperature should be a good approximation of $\Delta(\omega=0)$ at zero temperature. In simpler systems with no magnetic field and where $\chi$ is ignored, one defines $\Delta_0 = \Delta(\omega = \Delta_0)$, the value of the gap on the gap edge \cite{Schrieffer1964book, Vidberg1977AnalyticCont}. A similar, yet more complicated self-consistent definition could be used in our case. References \cite{EliashbergFlatBandSchrodi2020Mar, Aperis2015magneticField, Aperis2018MultibandEliashberg} indicate that $\Delta(\omega)$ varies slowly in a region around $\omega=0$ larger than the gap size. Hence, $\Delta_0 \approx \Delta(\omega = 0) \approx \Re\Delta(i\omega_{n=0})$ should be a good approximation. We also use $\Delta_0 = \Re\Delta(i\omega_{n=0})$ when plotting results as a function of temperature or magnetic field. However, when discussing the ratio $h_c/\Delta_0$ and the CC limit in this and the following two sections, $\Delta_0$ refers to our estimate of the zero temperature limit at zero magnetic field. 

In BCS theory, the dimensionless coupling $\lambda = V N_F$. Here, we let $NV = -g^2 D(i\omega_\nu = 0) = 2g^2/\omega_E$ to get an estimate of $\lambda$ in the strong coupling case. Also, it is known that $Z(i\omega_{n=0}) \approx 1 + \lambda$  \cite{EliashbergFlatBandOjajarvi2017}, given that FS averages provide reasonable approximations. In the case of one electron band, the two estimates agree very well, both giving $\lambda \approx 0.69$, with $tN_F/N \approx 0.14$ at $\mu = -t$ and $Z(i\omega_{n=0})$ shown in Fig.~\ref{fig:E1}. Additionally, $2\Delta_0/T_c \approx 4.0$ is greater than the BCS prediction $2\Delta_0/T_c \approx 3.5$, placing us in the strong-coupling limit \cite{SFsuperconductivity}.

\section{Two bands that cross} \label{sec:2band}
In order to compare our results with those in Ref.~\cite{Ghanbari2022Feb}, we take the same bands, one dispersive $\epsilon_{\boldsymbol{k},1} = \epsilon_{\boldsymbol{k},d} = -\mu-2t(\cos k_x + \cos k_y)$ and one completely flat $\epsilon_{\boldsymbol{k},2} = \epsilon_{f} = -\mu_0$. For the sake of generality, we do not specify the underlying microscopic model giving rise to the flat band. As noted previously, the form of EPC means that there will be hybridization between the bands. Here, we assume that hybridization effects are negligible. For multiband models, the DOS is $D(\epsilon) = \sum_{\boldsymbol{k}l\sigma} \delta(\epsilon-\epsilon_{\boldsymbol{k}l\sigma})$. With the second band being flat, this gives a delta function contribution to the DOS at the position of the flat band. Compared with Fig.~\ref{fig:DF1}(a) for the one band case, there is now a flat band just below the FS, with an accompanying spike in the DOS.

Reference \cite{Aase2023Constrained} considers the free energy of multiband superconductors using BCS theory and finds that in certain situations constraints must be placed on the gaps to ensure a free energy that is bounded from below. This analysis relies on the inverse of the matrix describing the intra- and interband couplings. The case we consider with all couplings equal, $V_{ij} = V$, where $i,j$ are band indices, is a special case where the $V_{ij}$ matrix is not invertible. However, with all couplings equal, $\Delta_i = \Delta$ for all band indices $i$ \cite{Ghanbari2022Feb}. In light of Ref.~\cite{Aase2023Constrained}, we view this as a constraint on the gaps. Then, the free energy becomes very similar to the one band case, containing a term $N\Delta^2/V$, and is obviously bounded from below \cite{Ghanbari2022Feb, Aase2023Constrained}. Reference \cite{Ghanbari2022Feb} considered both $V_{ij} = V$ and situations where the couplings had different strengths for different band indices. They found that the key to exceeding the CC limit was the intraband coupling in the flat band $V_{22}$ and that there was a not too small interband coupling $V_{12} = V_{21}$. For computational convenience, we stick to $V_{ij} = V$ in this paper.  

\begin{figure}
    \centering
    \includegraphics[width = \linewidth]{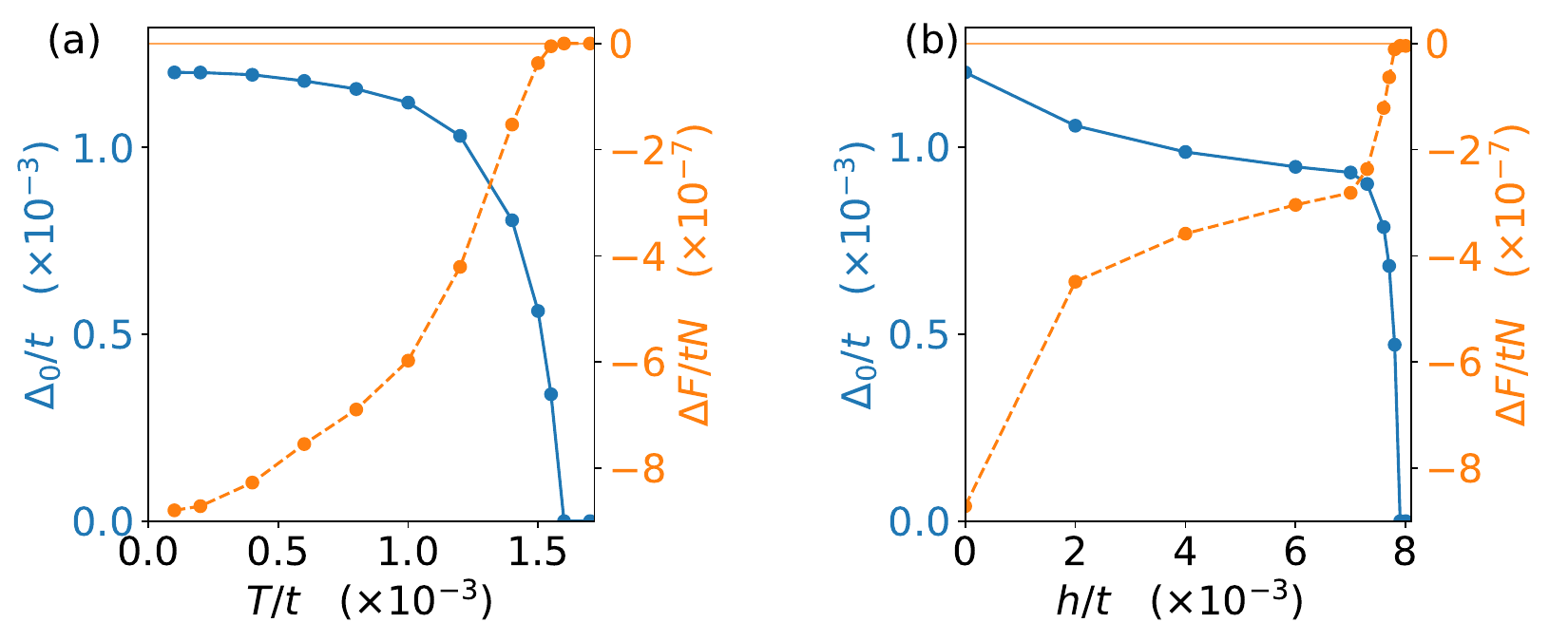}
    \caption{Results for a two-band system with one dispersive band and one flat band that cross. (a) The superconducting gap $\Delta_0$ (solid blue curve) as a function of temperature at $h=0$, giving $T_c/t \approx 1.6\times 10^{-3}$ and $\Delta_0/t \approx 1.2 \times 10^{-3}$. The free energy difference between the superconducting and normal state $\Delta F$ is shown as a dashed orange curve. (b) $\Delta_0$ as a function of magnetic field (solid blue curve), and $\Delta F$ (dashed orange curve) for $T/t = 2\times 10^{-4}$. $\Delta F$ stays negative as long as nonzero-gap solutions are found, giving $h_c/t \approx 7.9 \times 10^{-3}$ and $h_c/\Delta_0 \approx 6.6$. Parameters are $\mu = -t$, $\mu_0/t = 0.02$, $g/t = 0.03485$, $\omega_E/t = 0.1$, $M = 20\omega_E$, $N = 200^2$, and tolerance for convergence $10^{-5}$. }
    \label{fig:DF2}
\end{figure}

\begin{figure}
    \centering
    \includegraphics[width = \linewidth]{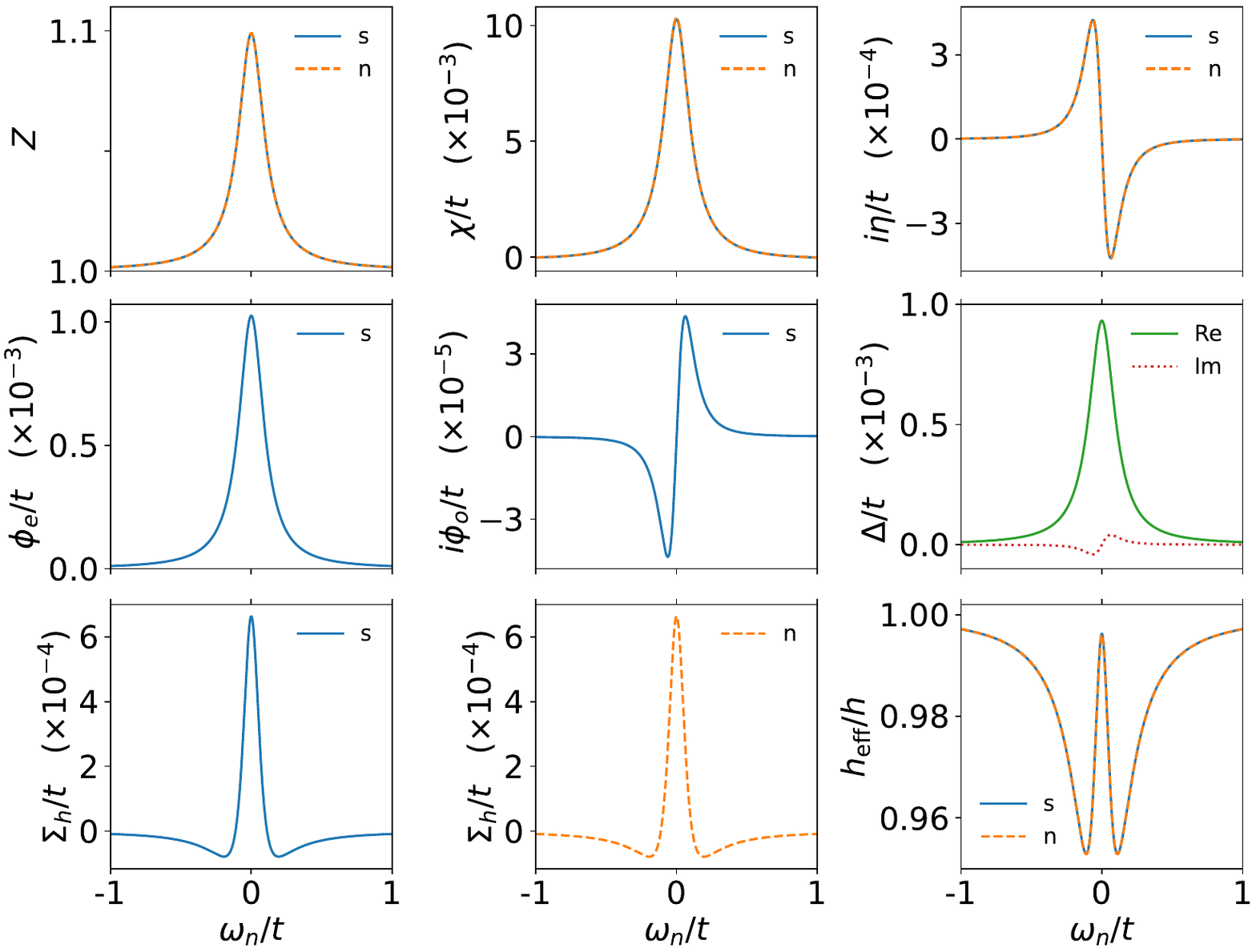}
    \caption{Solutions to the Eliashberg equations for the superconducting state (solid blue curves) and normal state (dashed orange curves) with two bands that intersect. The superconducting gap $\Delta(i\omega_n)$ is shown with the real part represented by a solid green curve and the imaginary part represented by a dotted red curve. We also show the effective magnetic field $h_{\text{eff}}(i\omega_n) = [h+\Sigma_h(i\omega_n)]/[Z(i\omega_n)]$ in the superconducting and normal states. Parameters are $h/t = 0.007$, $T/t = 2\times 10^{-4}$, and otherwise the same as in Fig.~\ref{fig:DF2}. }
    \label{fig:E2}
\end{figure}

Results for the case of a flat band placed just below the FS are shown in Fig.~\ref{fig:DF2}. We find $T_c/t \approx 1.6\times 10^{-3}$ and $\Delta_0/t \approx 1.2 \times 10^{-3}$ giving $2\Delta_0/T_c \approx 1.5$. A low value for $2\Delta_0/T_c$ was also found for the same system in Ref.~\cite{Ghanbari2022Feb} using a weak-coupling approach, so it appears that the flat band boosts $T_c$ more than it boosts the gap amplitude. Furthermore, we obtain $h_c/t \approx 7.9\times 10^{-3}$ giving $h_c/\Delta_0 \approx 6.6$, almost ten times greater than the CC limit.

Figure \ref{fig:E2} shows solutions to the Eliashberg equations with the same parameters as in Fig.~\ref{fig:DF2}. Since we now consider a smaller gap, the solutions for the normal state and superconducting state are more similar than in Fig.~\ref{fig:E1}. Compared with the one-band case in Fig.~\ref{fig:E1}, $\eta$ and $\chi$ have changed sign. We interpret this as a result of the fact that now, most of the states lie below the FS, while in Fig.~\ref{fig:E1} a majority of the states lie above the FS. Also note that $h_{\text{eff}}$ is much closer to $h$ now. Hence, exceeding the CC limit is definitely due to the flat band, not the renormalization of the magnetic field. 

Some fine tuning of parameters was involved in obtaining Fig.~\ref{fig:DF2}. Naturally, to get high $h_c/\Delta_0$, a low $\Delta_0$ helps. We tuned $g$ in order to get $\Delta_0/t \sim 0.001$. Renormalizations move the flat band closer to the FS, so the bare flat band is placed further down from the FS than in Ref.~\cite{Ghanbari2022Feb}. With $\max \chi(i\omega_n)/t \approx 0.01$ and $\mu_0/t = 0.02$ we would naively expect $h_c/t \approx 0.01$. The slightly lower result, $h_c/t \approx 0.008$, is caused by the model being more complicated than the simple picture that the normal state becomes energetically preferred once the flat band corresponding to spin down crosses the FS. Even in the BCS case, the situation is more complicated than that \cite{Ghanbari2022Feb}. Here, in the strong coupling approach, renormalizations in both the normal state and the superconducting state change as a function of the magnetic field, and the free energy is much more intricately linked with the magnetic field strength.

In the system we consider, $\Delta_0$ should be smaller than $\mu_0$ in order to get a large $h_c/\Delta_0$. Simultaneously, $\mu_0$ should be much smaller than $\omega_E$ to ensure small vertex corrections \cite{EliashbergFlatBandSchrodi2021Apr}. This places restrictions on parameter choices where the system we study here can exceed the CC limit by a large amount. One could imagine relaxing the restriction $\mu_0 \ll \omega_E$ by explicitly calculating and including vertex corrections. If the vertex corrections do not give dramatic changes to the results, one could then imagine increasing the EPC strength $g$, and so $\Delta_0$, and still achieving a large $h_c/\Delta_0$ by increasing $\mu_0$. An alternate view is that when $\Delta_0$ becomes very large, $h_c \approx \Delta_0$ is a very large magnetic field so that surpassing the CC limit becomes less important in applications. We leave the case of larger $\Delta_0$ and considerations of vertex corrections beyond the scope of this paper.

\section{Two hybridized bands} \label{sec:2bandhybrid}
\begin{figure}
    \centering
    \includegraphics[width = \linewidth]{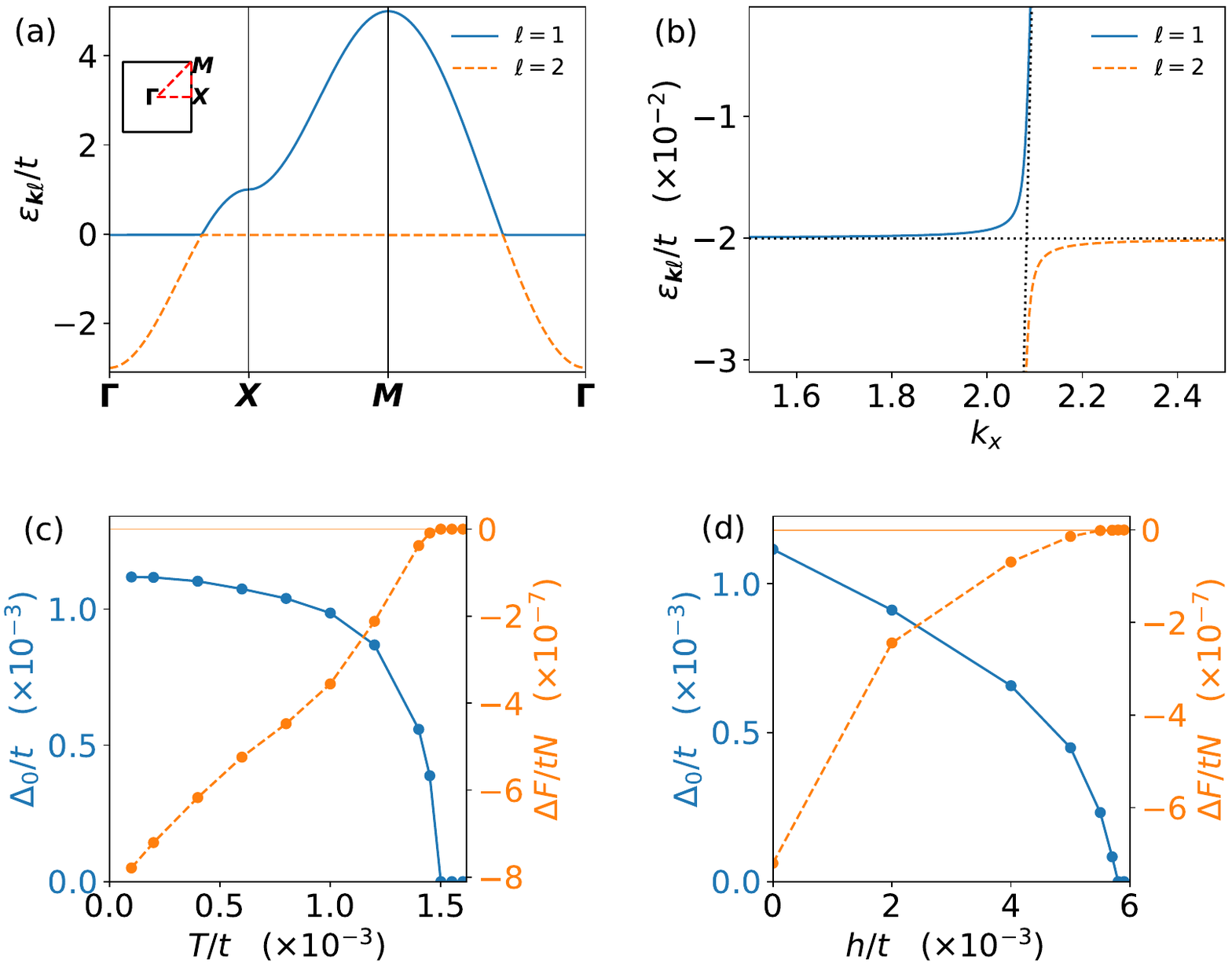}
    \caption{(a) The electron dispersion in Eq.~\eqref{eq:hybrid} for $h=0$, with a closer view of the hybridization given in (b) along with dotted black lines showing the original flat and dispersive bands. (c) The superconducting gap $\Delta_0$ (solid blue curve) as a function of temperature at $h=0$, giving $T_c/t \approx 1.5\times 10^{-3}$ and $\Delta_0/t \approx 1.1 \times 10^{-3}$. The free energy difference between the superconducting and normal states $\Delta F$ is shown as a dashed orange curve. (d) $\Delta_0$ as a function of magnetic field (solid blue curve), and $\Delta F$ (dashed orange curve) for $T/t = 2\times 10^{-4}$. $\Delta F$ stays negative as long as nonzero gap solutions are found, giving $h_c/t \approx 5.8 \times 10^{-3}$ and $h_c/\Delta_0 \approx 5.3$. Parameters are $\mu = -t$, $\mu_0/t = 0.02$, $\Sigma_{12}/t = 0.01$, $g/t = 0.03485$, $\omega_E/t = 0.1$, $M = 20\omega_E$, $N = 220^2$, and tolerance for convergence $10^{-5}$. }
    \label{fig:DF3}
\end{figure}

We now include hybridization between the two bands in the previous section. This originates with self-energy effects due to Feynman diagrams with different band indices in the external legs. 
We refer to these effects with a single frequency dependent function $\Sigma_{12}(i\omega_n)$. Since the EPC is chosen to be band independent, we expect that the functional form of $\Sigma_{12}(i\omega_n)$ is closely related to the intraband self-energies $\chi(i\omega_n)$, $\eta(i\omega_n)$ and $\Sigma_h(i\omega_n)$. The effect of the hybridization is strongest on the bands close to where they cross, which happens at a specific quasiparticle energy, so that only a small real-frequency region of $\Sigma_{12}(\omega)$ matters. We simplify, and take the hybridization parameter $\Sigma_{12}$ to be a real constant representing the average value in the most relevant region of frequencies. Furthermore, we ignore any spin dependent part of $\Sigma_{12}$. From the Dyson equation, the renormalized bands due to $\Sigma_{12}$ are the eigenvalues of 
\begin{equation}
    \begin{pmatrix}
        \epsilon_{\boldsymbol{k},d} & \Sigma_{12} \\
        \Sigma_{12} & \epsilon_{f}
    \end{pmatrix}.
\end{equation}
Hence the hybridized bands are
\begin{equation}
\label{eq:hybrid}
    \epsilon_{\boldsymbol{k}l} = \frac{1}
    {2}\pqty{\epsilon_{\boldsymbol{k},d}+\epsilon_{f} + n_l \sqrt{(\epsilon_{\boldsymbol{k},d}-\epsilon_{f})^2 + 4\Sigma_{12}^2}},
\end{equation}
with $n_1 = 1, n_2 = -1$. Figure \ref{fig:DF3}(a) shows this dispersion, while Fig.~\ref{fig:DF3}(b) highlights the hybridization. Now, the delta function peak in the DOS at $-\mu_0$ is replaced by zero DOS at $-\mu_0$ and a very large, finite DOS in small regions above and below $-\mu_0$. As previously noted, this band structure resembles that of the Anderson lattice model of Kondo insulators \cite{Dzero2016KondoInsulatorRev, Lai2018KondoSemimetal}, models of heavy-fermion superconductors \cite{Steglich2016Heavy-fermion}, and models for twisted-bilayer graphene \cite{Islam2023Feb}.

Figure \ref{fig:DF3} shows the temperature and magnetic field evolution of $\Delta_0$ and $\Delta F$ when including hybridization. We set $\Sigma_{12}$ approximately equal to the zero-imaginary-frequency limit of $\chi(i\omega_n)$. Then, we find  $T_c/t \approx 1.5\times 10^{-3}$, $\Delta_0/t \approx 1.1 \times 10^{-3}$, and $h_c/t \approx 5.8 \times 10^{-3}$. This yields $h_c/\Delta_0 \approx 5.3$, lower than when ignoring hybridization, but still far exceeding the CC limit. While hybridization reduces the effectiveness of the flat band, the large nearly flat regions still provide essentially the same effect as studied in Ref.~\cite{Ghanbari2022Feb}. The solutions to the Eliashberg equations are very similar to the case of zero hybridization in Fig.~\ref{fig:E2}, except that the gap amplitude decreases more rapidly with magnetic field. We checked whether this was an effect of being at an effectively higher temperature since $T_c$ is slightly smaller. Solutions at $T/t = 10^{-4}$, half the value used in Fig.~\ref{fig:DF3}(d), give essentially the same result for the magnetic field strengths we checked. At $h/t = 5\times 10^{-3}$, $\Delta_0/t \approx 0.45 \times 10^{-3}$ for both temperatures. The faster decrease of $\Delta_0(h)$ is therefore interpreted as being a result of the hybridization moving parts of the flat band region further from the FS. Also, at $T/t = 10^{-4}$, we found that $\Delta F$ changes sign before nonzero gap solutions disappear from the Eliashberg equations, confirming the low-temperature first-order phase transition. This slightly reduces the estimate of the critical magnetic field to $h_c/t = 5.7\times 10^{-3}$, giving $h_c/\Delta_0 \approx 5.2$, more than seven times the CC limit. While hybridization reduces $h_c/\Delta_0$, we expect that the system will continue to exceed the CC limit by a large amount if the strength of the hybridization is increased.

Our strong-coupling consideration essentially shows that the results derived using BCS theory in Ref.~\cite{Ghanbari2022Feb} are valid if the bands used there are viewed as already renormalized. Aside from hybridization, renormalization effects do not add dispersion to the flat band, so the underlying mechanism discussed in Ref.~\cite{Ghanbari2022Feb} also applies at strong coupling. Renormalizations do shift the position of the flat band though, which is a crucial parameter when it comes to exceeding the CC limit. It is also predicted that renormalization effects in flat band systems give rise to replica bands appearing at integer multiples of the Einstein frequency in the spectral function \cite{EliashbergFlatBandSchrodi2021Apr}. Since this is far from the FS with the parameters in Figs.~\ref{fig:DF2} and \ref{fig:DF3}, we do not expect that they play a significant role in our system. Future studies could apply our setup to a specific material with a microscopic model in mind. In that case, a detailed study of the quantum geometry is possible, to ensure a nonzero superfluid weight. It is likely that the flat band should not be a trivial insulating band originating with electrons that are localized at separate lattice sites. Rather, a nontrivial flat band should be pursued \cite{PeottaTorma2015FlatBand, Torma2022QuantumGeom}. However, the hybridization with a dispersive band may still give a nonzero superfluid weight of the two-band system even with a trivial flat band.

\section{Conclusion} \label{sec:conclusion}
We have studied the phase transition from a superconducting state to a normal state caused by the application of an in-plane magnetic field in a 2D superconductor. The Chandrasekhar-Clogston limit puts an upper limit on the magnetic field at $h_c/\Delta_0 \approx 0.7$ for conventional weak-coupling superconductors. Using strong-coupling theory, we have derived Eliashberg equations for a multiband system in a magnetic field. Following a Green's function approach, superconductivity and self-energy effects are treated in a self-consistent manner. We have derived the free energy using functional integral methods and expressed it in terms of solutions to the Eliashberg equations.

Applied to a one-band strong-coupling system, we found that renormalization of the magnetic field caused by self-energy effects can lead to a slightly larger critical field than the CC limit. We have also studied a two-band system with a flat band placed just below the Fermi level of a dispersive band. Weak-coupling calculations show that such a system can exceed the CC limit. The large density of states for the flat band suggests the presence of strong coupling, so that strong-coupling calculations may offer corrections to weak-coupling predictions. Nevertheless, we have found that such a system can surpass the CC limit by a factor of almost 10, also when using strong-coupling theory. Finally, we have considered the presence of self-energy-driven hybridization between the bands. While the critical magnetic field decreases, the system still exceeds the CC limit by a factor of 7.

\section*{Acknowledgments}
We thank Niels Henrik Aase and Jacob Linder for helpful comments on the manuscript. 
We acknowledge funding from the Research Council of Norway (RCN) through its Centres of Excellence funding scheme Project No.~262633, ``QuSpin," and RCN through Project No.~323766, ``Equilibrium and out-of-equilibrium quantum phenomena in superconducting hybrids with antiferromagnets and topological insulators."

\appendix

\section{Full Eliashberg equations} \label{app:fullEliashberg}
For completeness, we state the full Eliashberg equations for a multiband system in a magnetic field with band-independent EPC. They are given by
\begingroup
\allowdisplaybreaks
\begin{align}
    Z(k) =& 1-\frac{1}{2 i\omega_{n}}\sum_{k'l} \lambda_{kk'} \bqty{\frac{i\omega_{n'} Z(k')}{\Theta_{l}^{+}(k')} + \frac{h + \Sigma_h(k')}{\Theta_{l}^{-}(k')}}, \\
    \Sigma_h(k) =& \frac{-1}{2}\sum_{k'l} \lambda_{kk'}\bqty{\frac{i\omega_{n'} Z(k')}{\Theta_{l}^{-}(k')} + \frac{h + \Sigma_h(k')}{\Theta_{l}^{+}(k')}}, \\
    \chi(k) =& \frac{1}{2}\sum_{k'l} \lambda_{kk'}\bqty{\frac{\epsilon_{\boldsymbol{k}'l}+\chi(k')}{\Theta_{l}^{+}(k')} +\frac{\eta(k')}{\Theta_{l}^{-}(k')} }, \\
    \eta(k) =& \frac{1}{2}\sum_{k'l} \lambda_{kk'}\bqty{\frac{\epsilon_{\boldsymbol{k}'l}+\chi(k')}{\Theta_{l}^{-}(k')} +\frac{\eta(k')}{\Theta_{l}^{+}(k')} }, \\
    \phi_e^R(k) =& \frac{-1}{2}\sum_{k'l} \lambda_{kk'}\bqty{\frac{\phi_e^R (k')}{\Theta_{l}^{+}(k')} - \frac{\phi_o^R (k')}{\Theta_{l}^{-}(k')}}, \\
    \phi_e^I(k) =& \frac{-1}{2}\sum_{k'l} \lambda_{kk'}\bqty{\frac{\phi_e^I (k')}{\Theta_{l}^{+}(k')} - \frac{\phi_o^I (k')}{\Theta_{l}^{-}(k')}}, \\
    \phi_o^R(k) =& \frac{-1}{2}\sum_{k'l} \lambda_{kk'}\bqty{-\frac{\phi_e^R(k')}{\Theta_{l}^{-}(k')} + \frac{\phi_o^R(k')}{\Theta_{l}^{+}(k')}},\\
    \phi_o^I(k) =& \frac{-1}{2}\sum_{k'l} \lambda_{kk'}\bqty{-\frac{\phi_e^I(k')}{\Theta_{l}^{-}(k')} + \frac{\phi_o^I(k')}{\Theta_{l}^{+}(k')}},
\end{align}
\endgroup
where $\lambda_{kk'} = -|g_{\boldsymbol{k}-\boldsymbol{k}'}|^2 D(k-k')$.

\section{Functional integral approach and free energy} \label{app:funkint}
Reference \cite{Protter2021funkintFE} introduces a way to derive the Eliashberg equations using functional integral methods \cite{Altland2010}. This paves the way for studying fluctuations around the superconducting state in a strong-coupling approach. In the present setting, however, its main interest is that it provides a way to derive a generalizable expression for the free energy, given solutions to the Eliashberg equations. We follow the derivation of Ref.~\cite{Protter2021funkintFE} but generalize to multiband systems in the presence of a magnetic field.

The starting point is to write the Hamiltonian in Eq.~\eqref{eq:Hamiltonian} as an action,
\begin{align}
    S[\bar{c}, c, \bar{a}, a] =& \int_0^\beta d\tau \bigg[\sum_{\boldsymbol{k}l\sigma} \bar{c}_{\boldsymbol{k}l\sigma}(\partial_\tau + \epsilon_{\boldsymbol{k}l}-\sigma h)  {c}_{\boldsymbol{k}l\sigma} \nonumber\\
    &+ \sum_{\boldsymbol{q}} \bar{a}_{\boldsymbol{q}} (\partial_\tau +\omega_{\boldsymbol{q}}) a_{\boldsymbol{q}} \nonumber\\
    &+\sum_{\boldsymbol{k}\boldsymbol{q}\sigma ll'} g_{\boldsymbol{q}} \bar{c}_{\boldsymbol{k}+\boldsymbol{q},l',\sigma } c_{\boldsymbol{k}l\sigma }(a_{\boldsymbol{q}}+\bar{a}_{-\boldsymbol{q}})  \bigg],
\end{align}
where fermion operators are replaced by Grassmann fields, and boson operators are replaced by complex fields. The fields depend on imaginary time, so ${c}_{\boldsymbol{k}l\sigma} = {c}_{\boldsymbol{k}l\sigma}(\tau)$, $a_{\boldsymbol{q}} = a_{\boldsymbol{q}}(\tau)$ and so on. Inserting the FTs
\begin{equation}
    {c}_{\boldsymbol{k}l\sigma}(\tau) = \frac{1}{\beta} \sum_{\omega_n} {c}_{kl\sigma}e^{-i\omega_n \tau}, {a}_{\boldsymbol{q}}(\tau) = \frac{1}{\beta} \sum_{\omega_\nu} {a}_{q}e^{-i\omega_\nu \tau},
\end{equation}
gives
\begin{align}
    S[\bar{c}, c, \bar{a}, a] =&  -\sum_{kl\sigma} \bar{c}_{kl\sigma}G_{0l\sigma}^{-1}(k)  {c}_{kl\sigma} - \sum_{q} \bar{a}_{q} \tilde{D}_0^{-1}(q) a_{q} \nonumber\\
    &+\sum_{kq\sigma ll'} g_{\boldsymbol{q}} \bar{c}_{k+q,l',\sigma } c_{k,l,\sigma }(a_{q}+\bar{a}_{-q})  ,
\end{align}
with $G_{0l\sigma}^{-1}(k) = i\omega_n - \epsilon_{\boldsymbol{k}l}+\sigma h$ and $\tilde{D}_0^{-1} (q) = i\omega_\nu-\omega_{\boldsymbol{q}}$. These are bare Greens functions for ${c}_{kl\sigma}$ and $a_{q}$ respectively \cite{Maeland2021Sep, BruusFlensberg}. 
The grand canonical partition function is
\begin{equation}
    Z_g = \int \mathcal{D}[\bar{c}, c, \bar{a}, a]e^{-S[\bar{c}, c, \bar{a}, a]},
\end{equation}
where $\mathcal{D}[\bar{c}, c, \bar{a}, a]$ is the path integral measure \cite{Protter2021funkintFE, Altland2010}.

\subsection{Integrate out phonons} 
The fermionic density is defined as $\rho_{q} = \sum_{k\sigma ll'} \bar{c}_{k+q,l',\sigma}c_{kl\sigma}$, with the property $\rho_q = \bar{\rho}_{-q}$. Factoring out the phonon part of $Z_g$ yields
\begin{align}
    Z_{\text{ph}}[\bar{c},c] =& \int \mathcal{D}[\bar{a}, a]\exp\bigg[\sum_q \big(\bar{a}_{q} \tilde{D}_0^{-1} a_{q} \nonumber\\
    &-g_{\boldsymbol{q}} (a_{q}+\bar{a}_{-q})\rho_q\big)\bigg].
\end{align}
The Gaussian integral can be computed by completing squares. Let $\bar{a}_q \to \bar{a}_q + g_{\boldsymbol{q}}\tilde{D}_0 (q)\bar{\rho}_{-q}$ and $a_q \to a_q + g_{-\boldsymbol{q}}\tilde{D}_0(q) \rho_{-q}$. Using that $\rho_q$ and $\bar{\rho}_q$ commute since the Grassmann variables anticommute, and that $g_{-\boldsymbol{q}} = g_{\boldsymbol{q}}^*$ from Hermiticity, we get
\begin{equation}
    Z_{\text{ph}}[\bar{c},c] = \mathcal{N}\exp\pqty{-\sum_q |g_{\boldsymbol{q}}|^2 \tilde{D}_0(q) \rho_q \bar{\rho}_q }.
\end{equation}
$\mathcal{N} = \exp \big[-\beta\sum_q \tr \ln \beta\tilde{D}_0^{-1}(q)\big]$ represents a phonon dependent contribution to the free energy. We ignore any feedback effects on the phonons and assume that their contribution is the same in the normal and superconducting states. Hence, $\mathcal{N}$ is an unimportant constant shifting the free energy. Note that $\tilde{D}_0(q)$ is coupled to terms that are even in $q$. As a result, only the part that is even in $q$ contributes. This is the propagator $D_0(q) \equiv \tilde{D}_0(q) + \tilde{D}_0(-q)$ of $a_{q}+\bar{a}_{-q}$. So, we write
\begin{equation}
    Z_{\text{ph}}[\bar{c},c] = \exp\pqty{-\frac{1}{2}\sum_q |g_{\boldsymbol{q}}|^2 D_0(q) \rho_q \bar{\rho}_q }.
\end{equation}

The full partition function is now
\begin{align}
    Z_g =& \int \mathcal{D}[\bar{c}, c]\exp\bigg[\sum_{kl\sigma} \bar{c}_{kl\sigma}G_{0l\sigma}^{-1}(k)  {c}_{kl\sigma} \nonumber \\
    &+ \frac{1}{2}\sum_q \lambda(q)\rho_q \bar{\rho}_q \bigg],
\end{align}
where $\lambda(q) = -|g_{\boldsymbol{q}}|^2 D_0(q)$. At this point, Ref.~\cite{Protter2021funkintFE} proceeds to FT to real space, perform a Hubbard-Stratonovich (HS) decoupling, and then FT back to momentum space. In the following, we perform the HS decoupling in momentum space, obtaining a generalization of the same end result.

\subsection{HS decoupling}
The full interaction term is
\begin{widetext}
\begin{equation}
    \frac{1}{2}\sum_q \lambda(q)\rho_q \bar{\rho}_q = \frac{1}{2}\sum_{kk'q\sigma\sigma'} \sum_{ll'l''l'''} \lambda(q)\bar{c}_{k+q,l',\sigma} c_{kl\sigma} \bar{c}_{k'l'''\sigma'} c_{k'+q,l'',\sigma'},
\end{equation}
which is too complicated to perform an HS decoupling, as there can only be two independent momentum sums. We also make certain assumptions on spin and band combinations that are typical for density-density type terms and BCS-type terms in multiband SCs. Consider the following parts of the interaction:
\begin{equation}
    \frac{1}{2 }\sum_{kk'q \sigma ll'} \lambda(q) \pqty{\bar{c}_{k+q,l',\sigma} c_{kl\sigma} \bar{c}_{k'l\sigma} c_{k'+q,l',\sigma} + \bar{c}_{k+q,l',\sigma} c_{kl\sigma} \bar{c}_{k',l',-\sigma} c_{k'+q,l,-\sigma}}.
\end{equation}
The first term is decoupled in the density-density channel, and the second is decoupled in the Cooper channel. We restrict ourselves to two momentum indices, guided by the physics we want to describe. In the first term, we choose $k' = k$, and then redefine $k+q = k'$, such that $q = k'-k$. That way, we get a quite generic density-density interaction in momentum space. In the second term, we assume opposite momentum pairing, first setting $q = -k'-k$ and then letting $k'\to -k'$. We get
\begin{equation}
    \frac{1}{\beta }\sum_{kk' ll'} \lambda(k-k') \pqty{-\frac{1}{2}\sum_\sigma  \bar{c}_{kl\sigma} c_{kl\sigma} \bar{c}_{k'l'\sigma} c_{k'l'\sigma} + \bar{c}_{k'l'\uparrow}\bar{c}_{-k',l',\downarrow}c_{-k,l,\downarrow}c_{kl\uparrow}}.
\end{equation}
Here, $1/\beta$ comes from removing one sum over momentum, where we have now picked out only one term. An argument for the above restriction on $q$ in the Cooper channel is that we are considering only EPC here. In reality, there would be a Coulomb interaction, which the EPC is only able to overcome for a few special choices of spin and momentum indices.

To perform the HS decoupling, we introduce auxiliary bosonic fields $\bar{\phi}, \phi, \Sigma$ and a measure $\mathcal{D}[\bar{\phi}, \phi, \Sigma]$ chosen such that
\begin{equation}
    1 = \int \mathcal{D}[\bar{\phi}, \phi, \Sigma] \exp \pqty{ -\frac{\beta}{N}\sum_{kk'}\lambda^{-1}(k-k') \bqty{\bar{\phi}(k) \phi(k') + \frac{1}{2}\sum_\sigma \Sigma^\sigma(k) \Sigma^\sigma (k')}}.
\end{equation}
Here, $\lambda^{-1}(q)$ is the FT of $1/\lambda(\boldsymbol{r}_i, \tau)$. The FT back to real space and imaginary time is defined as
\begin{equation}
    \lambda(\boldsymbol{r}_i, \tau) = \frac{1}{\sqrt{N}}\sum_{\boldsymbol{q}}\frac{1}{\beta}\sum_{i\omega_\nu} \lambda (\boldsymbol{q}, i\omega_\nu) e^{i(\boldsymbol{q}\cdot \boldsymbol{r}_i -\omega_\nu \tau)}.
\end{equation}
Meanwhile, $\lambda^{-1}(q)$ is defined as
\begin{equation}
    \lambda^{-1}(\boldsymbol{q}, i\omega_\nu) = \frac{1}{\sqrt{N}}\sum_{\boldsymbol{r}_i}\int_0^\beta d\tau \frac{1}{\lambda(\boldsymbol{r}_i, \tau)}e^{-i(\boldsymbol{q}\cdot \boldsymbol{r}_i -\omega_\nu \tau)}.
\end{equation}
$\lambda^{-1}(q)$ defined as the FT of $1/\lambda(\boldsymbol{r}_i, \tau)$ satisfies
\begin{equation}
\label{eq:laminverse}
    \sum_{k''} \lambda(k-k'') \lambda^{-1}(k''-k') = N\beta \delta_{\boldsymbol{k}\boldsymbol{k}'}\delta_{i\omega_n,i\omega_{n'}} = N\beta \delta_{k,k'}.
\end{equation}

Now,
\begin{equation}
    Z_g = \int \mathcal{D}[\bar{c}, c, \bar{\phi}, \phi, \Sigma]\exp\pqty{-S_F[\bar{c},c]-\frac{\beta}{N}\sum_{kk'} \lambda^{-1}(k-k') \bqty{\bar{\phi}(k) \phi(k') + \frac{1}{2}\sum_\sigma \Sigma^\sigma(k) \Sigma^\sigma (k')}},
\end{equation}
\begin{equation}
    S_F[\bar{c},c] = -\sum_{kl\sigma} \bar{c}_{kl\sigma}G_{0l\sigma}^{-1}(k)  {c}_{kl\sigma}-\frac{1}{\beta }\sum_{kk' ll'} \lambda(k-k') \pqty{-\frac{1}{2}\sum_\sigma  \bar{c}_{kl\sigma} c_{kl\sigma} \bar{c}_{k'l'\sigma} c_{k'l'\sigma} + \bar{c}_{k'l'\uparrow}\bar{c}_{-k',l',\downarrow}c_{-k,l,\downarrow}c_{kl\uparrow}}.
\end{equation}
We treat $\Sigma^\sigma$ as density fluctuations and $\phi$ as fermion pairing. To eliminate the four-fermion terms, we introduce the shifts
\begin{align}
    \bar{\phi}(k) &\to \bar{\phi}(k)-\frac{1}{\beta}\sum_{k''l}\lambda(k-k'') \bar{c}_{k''l\uparrow}\bar{c}_{-k'',l,\downarrow}, \\
    \phi(k') &\to \phi(k')-\frac{1}{\beta} \sum_{k''l} \lambda(k'-k'') c_{-k'',l,\downarrow}c_{k''l\uparrow}, \\
    \Sigma^\sigma(k) &\to \Sigma^\sigma (k) + \frac{i}{\beta}\sum_{k''l}\lambda(k-k'') \bar{c}_{k''l\sigma}c_{k''l\sigma} .
\end{align}
Using Eq.~\eqref{eq:laminverse}, we get
\begin{align}
    S[\bar{c}, c, \bar{\phi}, \phi, \Sigma] =& \frac{\beta}{N}\sum_{kk'} \lambda^{-1}(k-k')\bqty{\bar{\phi}(k) \phi(k') + \frac{1}{2} \Sigma^\sigma(k) \Sigma^\sigma (k')}\nonumber \\
    & - \sum_{kl} \pqty{\bar{c}_{kl\sigma}[G_{0l\sigma}^{-1}(k)-i\Sigma^\sigma(k)]c_{kl\sigma} + \phi(k)\bar{c}_{kl\uparrow}\bar{c}_{-k,l,\downarrow} + \bar{\phi}(k)c_{-k,l,\downarrow}c_{kl\uparrow} }.
\end{align}
Sums over repeated $\sigma$ indices are left implicit.

\subsection{Integrate out electrons}
With Nambu spinor $\bar{\Psi}_{kl} = (\Bar{c}_{kl\uparrow}, c_{-k,l,\downarrow})$,
\begin{align}
    S[\bar{c}, c, \bar{\phi}, \phi, \Sigma] =& \frac{\beta}{N}\sum_{kk'} \lambda^{-1}(k-k') \bqty{\bar{\phi}(k) \phi(k') + \frac{1}{2} \Sigma^\sigma(k) \Sigma^\sigma (k')} - \sum_{kl} \Bar{\Psi}_{kl} \Tilde{G}_l^{-1}(k) \Psi_{kl},
\end{align}
where
\begin{equation}
    \tilde{G}_l^{-1}(k) = \begin{pmatrix}    G_{0l\uparrow}^{-1}(k)-i\Sigma^\uparrow (k) & \phi (k) \\
    \bar{\phi}(k) & -G_{0l\downarrow}^{-1}(-k) +i\Sigma^\downarrow(-k)
    \end{pmatrix}.
\end{equation}
Integrating out the electrons then gives the action
\begin{equation}
    S_{\text{HS}}[\bar{\phi}, \phi, \Sigma] = -\beta \sum_{kl} \tr \ln (-\beta \tilde{G}_l^{-1}(k)) +\frac{\beta}{N}\sum_{kk'} \lambda^{-1}(k-k') \bqty{\bar{\phi}(k) \phi(k') + \frac{1}{2} \Sigma^\sigma(k) \Sigma^\sigma (k')}.
\end{equation}

\subsection{Eliashberg equations and free energy}
The grand canonical partition function $Z_g = \int \mathcal{D}[\bar{\phi}, \phi, \Sigma]\exp (-S_{\text{HS}}[\bar{\phi}, \phi, \Sigma]) = e^{-\beta \Omega}$, where $\Omega = F-\mu N$ is the grand potential and $F$ is the free energy. The main contribution to $Z_g$ is expected to come from regions close to the stationary points of the action $S_{\text{HS}}$ \cite{Protter2021funkintFE}, where it is slowly varying. Following the same route as Ref.~\cite{Protter2021funkintFE}, the stationary point conditions are
\begin{align}
    i\Sigma^\sigma(k) =& \delta_{\sigma\uparrow} \sum_{k' l}\lambda(k-k') \frac{-G_{0l\downarrow}^{-1}(-k')+i\Sigma^\downarrow (-k')}{\Phi_{1l}(k')} +\delta_{\sigma\downarrow} \sum_{k' l}\lambda(k-k') \frac{-G_{0l\uparrow}^{-1}(-k')+i\Sigma^\uparrow (-k')}{\Phi_{2l}(k')}, \label{eq:statpt1} \\
    \phi(k) =& -\sum_{k'l} \lambda(k-k') \frac{\phi(k')}{\Phi_{1l}(k')}, \qquad \bar{\phi}(k) = -\sum_{k'l} \lambda(k-k') \frac{\bar{\phi}(k')}{\Phi_{1l}(k')}, \label{eq:statpt3}
\end{align}
where $\Phi_{1l}(k) = \det \tilde{G}_l^{-1}(k) = [G_{0l\uparrow}^{-1}(k)-i\Sigma^\uparrow (i\omega_{n})][-G_{0l\downarrow}^{-1}(-k)+i\Sigma^\downarrow(-i\omega_{n})]-\phi(i\omega_{n})\bar{\phi}(i\omega_{n})$ and $\Phi_{2l}(k) = \Phi_{1l}(-k)$.
Upon comparing the $G^{-1}$ matrices in the Green's function approach and the functional integral approach, we identify
\begin{align}
    i\Sigma^\uparrow (k) &= i\omega_n - Z(k)i\omega_n +\chi(k)+\eta(k)-\Sigma_h(k), \\
    i\Sigma^\downarrow (k) &= i\omega_n -Z(k)i\omega_n +\chi(k)-\eta(k)+\Sigma_h(k), \\
    \phi(k) = \bar{\phi}(k) &= \phi_e(k)-\phi_o(k).
\end{align}
Then, $\Phi_{1l} = \Theta_{1l}, \Phi_{2l} = \Theta_{2l}$. Using symmetries, the stationary point conditions can be rewritten as the Eliashberg equations given in Appendix \ref{app:fullEliashberg}.

$\lambda^{-1}(q)$ is an oscillating function of frequency \cite{Aase2023Eliashberg}. When $T \ll \omega_E$, its magnitude becomes very large, meaning that any calculation involving $\lambda^{-1}(q)$ and a sum over frequencies becomes a numerical sign problem. It is convenient to insert the stationary point conditions in the action to eliminate $\lambda^{-1}(q)$.
Inserting Eqs.~\eqref{eq:statpt1} and \eqref{eq:statpt3} in $S_{\text{HS}}$ gives the variational action
\begin{align}
    S_{\text{var}}[\bar{\phi}, \phi, \Sigma] =& -\beta \sum_{kl} \tr \ln (-\beta \tilde{G}_l^{-1}(k)) +\beta\sum_{kk' ll'}\lambda(k-k') \nonumber \\
    &\times \bigg[ \frac{\Bar{\phi}(k)\phi(k')}{\Phi_{1l}(k)\Phi_{1l'}(k')}-\frac{1}{2}\frac{[G_{0l\downarrow}^{-1}(-k)-i\Sigma^\downarrow(-k)][G_{0l'\downarrow}^{-1}(-k')-i\Sigma^\downarrow(-k')]}{\Phi_{1l}(k)\Phi_{1l'}(k')} \nonumber \\
    &-\frac{1}{2}\frac{[G_{0l\uparrow}^{-1}(-k)-i\Sigma^\uparrow(-k)][G_{0l'\uparrow}^{-1}(-k')-i\Sigma^\uparrow(-k')]}{\Phi_{2l}(k)\Phi_{2l'}(k')}\bigg].
\end{align}
The name ``variational action'' comes from the fact that inserting the stationary point conditions in the action yields an expression similar to the free energy derived in a variational approach \cite{Chubukov2022FEgeneral}. From $Z_g = e^{-\beta \Omega} = \int \mathcal{D}[\bar{\phi}, \phi, \Sigma]\exp (-S_{\text{HS}}[\bar{\phi}, \phi, \Sigma]) \approx e^{-S_{\text{var}}[\bar{\phi}, \phi, \Sigma]}$ under the assumption that the main contribution comes from the stationary point \cite{Protter2021funkintFE}, the grand potential is $\Omega \approx S_{\text{var}}[\bar{\phi}, \phi, \Sigma]/\beta$ \cite{Lundemo2023, Aase2023Eliashberg, Syljuasen2021FreeEnergy}. We assume that the chemical potential is unchanged between the normal and superconducting states, as discussed in Refs.~\cite{Bardeen1964ElishbergFreeEnergyT0, Wada1964ElishbergFreeEnergyT0}, so that $\Delta F = \Delta \Omega = F_s-F_n$ provides an expression for the free energy difference between the superconducting and normal states. Specializing to Einstein phonons with isotropic EPC, and using $\tr \ln A = \ln \det A$ yields the free energy in Eq.~\eqref{eq:FE}. Note that the definition of the momentum independent $\lambda^{-1}(i\omega_\nu)$ is changed to
\begin{equation}
    \frac{1}{\beta} \sum_{i\omega_{n''}} \lambda(i\omega_{n}-i\omega_{n''})\lambda^{-1}(i\omega_{n''}-i\omega_{n'}) = \beta \delta_{i\omega_{n},i\omega_{n'}},
\end{equation}
compared with Eq.~\eqref{eq:laminverse}. $\lambda^{-1}(i\omega_\nu) = \int_0^\beta d\tau e^{i\omega_\nu \tau}/\lambda(\tau)$ satisfies this.
All results in this general appendix carry over to the specific case of momentum independent $\lambda(i\omega_\nu)$ quite straightforwardly. With the aid of Ref.~\cite{Aase2023Eliashberg}, generalizing the calculations in this appendix to the case of band-dependent interactions and, consequently, band-dependent Eliashberg functions is straightforward.

\end{widetext}

\bibliography{main.bbl}

\end{document}